\documentclass[12pt]{article}
\usepackage[onehalfspacing]{setspace}
\usepackage[margin=0.8in]{geometry}
\usepackage{amsmath,amsfonts,amssymb,mathrsfs}
\usepackage{ntheorem}
\usepackage{bbm}
\usepackage{graphicx}
\usepackage{enumerate}
\usepackage{xcolor}
\usepackage[round]{natbib}
\usepackage[hidelinks,colorlinks=true,linkcolor=blue,citecolor=blue]{hyperref}
\usepackage{appendix}
\usepackage{rotating}
\usepackage{ifthen}

\renewcommand{\P}{\mathbb{P}}
\newcommand{\E}{\mathbb{E}}
\newcommand{\V}{\mathbb{V}}
\newcommand{\cov}{\mathbb{C}\mathrm{ov}}
\newcommand{\1}{\mathbbm{1}}

\newcommand{\abs}[1]{\left\vert#1\right\vert}
\newcommand{\norm}[1]{\left\lVert#1\right\rVert}

\newcommand{\X}{\mathbf{X}}
\newcommand{\tstar}{\mathbf{t}^*}

\newtheorem{theorem}{Theorem}
\newtheorem{lemma}{Lemma}
\newtheorem{applemma}{Lemma}
\newtheorem{proposition}{Proposition}

\newtheorem{assumption}{Assumption}
\newtheorem{remark}{Remark}

\title{Post-Matching Two-Way Fixed Effects Estimation\thanks{We thank Max Farrell and Doug Steigerwald for valuable comments and suggestions.}}
\author{Yihong Liu\thanks{Department of Economics, UC Santa Barbara.} {and} Gonzalo Vazquez-Bare\thanks{Department of Economics, UC Santa Barbara.}}
\date{\today}

\begin{document}

\maketitle

\begin{abstract}
When estimating treatment effects with two-way fixed effects (2WFE) models, researchers often use matching as a pre-processing step when the parallel trends assumption is thought to hold conditionally on covariates. Specifically, in a first step, each treated unit is matched to one or more untreated units based on observed time-invariant covariates. In the second step, treatment effects are estimated with a 2WFE regression in the matched sample, reweighting the untreated units by the number of times they are matched. We formally analyze this common practice and highlight two problems. First, when different treatment cohorts enter treatment in different time periods, the post-matching 2WFE estimator that pools all treated cohorts has an asymptotic bias, even when the treatment effect is constant across units and over time. Second, failing to account for the variability introduced by the matching procedure yields invalid standard error estimators, which can be biased upwards or downwards depending on the data generating process. We propose simple post-matching difference-in-differences estimators that compare each treated cohort to the never-treated separately, instead of pooling all treated cohorts. We provide conditions under which these estimators are consistent for well-defined causal parameters, and derive valid standard errors that account for the matching step. We illustrate our results with simulations and with an empirical application. 
\end{abstract}

\vspace{2em}


\newpage



\section{Introduction}

When estimating treatment effects with two-ways fixed effects (2WFE) regressions, researchers often use matching as a pre-processing step. In such settings, each treated unit is matched to an untreated unit using pre-treatment observed covariates. In a second step, a 2WFE specification is estimated in the matched sample, reweighting the untreated units by the number of times they are used as a match for the treated units. This approach is typically justified under a conditional parallel trends assumption, by which units that are similar in their observed covariates are more likely to have followed the same outcome trends in the absence of treatment. While several alternative methods are available for estimating treatment effects under a conditional parallel trends assumption, matching-based methods have the advantage of being intuitively appealing, easy to implement, computationally stable and fully nonparametric, not requiring the functional form assumptions often invoked in inverse-probability weighting or regression adjustment methods. This combination of matching and 2WFE is common in empirical work: Table \ref{tab:matching_ddd_literature} lists 22 papers published between 2013 and 2025 in top general interest and field journals in Economics and Political Science that use some version of this approach.

In this paper, we formally analyze this commonly employed strategy and highlight two problems that generally invalidate its findings. First, when different treatment cohorts enter treatment in different time periods, the post-matching 2WFE estimator that pools all treated cohorts converges to a linear combination of average treatment effects on the treated (ATTs) plus an asymptotic bias, which we refer to as ``pooling bias'', even when treatment effects are constant across units and over time. To explain this result, we show that the post-matching 2WFE estimator can be decomposed into a linear combination of 2x2 (that is, two-group-two-period) reweighted difference-in-differences (DiD) estimators. These 2x2 estimators compare  each treated cohort to the never-treated cohort, and each treatment cohort to other treatment cohorts, with a specific set of weights. Because each treated unit is matched to a never-treated unit in the first step, comparisons between treated and reweighted never-treated units consistently estimate ATTs for different cohorts at different time periods. When comparing units across different treatment cohorts, however, units are not matched to each other, and thus the group used as comparison is not reweighted. As a result, the two groups will generally exhibit different covariate distributions. When the parallel trends assumption holds conditionally, these unweighted comparisons between different treated cohorts are invalid, even when the comparison group consists of not-yet-treated units. 

To address this issue, we propose post-matching difference-in-differences (DiD) estimators that compare each treated cohort to the never-treated units separately, instead of pooling all treated cohorts. We show that under the conditional parallel trends assumption and mild regularity conditions, these pairwise estimators are consistent for well-defined averages of ATTs over time. 

The second problem is that matching introduces variability that needs to be accounted for when conducting inference, and failing to account for this additional variability results in invalid standard errors. We characterize the probability limit of the naive cluster-robust variance estimator for the post-matching DiD estimator and show that failing to account for the matching step creates an asymptotic bias in the variance estimator, which can be positive or negative depending on the data generating process. We rely on recent advancements on measures of high-order Voronoi cells by \citet{Chen-Han_2024_wp} to provide a closed-form representation of the correct limiting variance of our proposed estimators, a result previously unavailable in the literature, and propose consistent variance estimators. Based on these results, we also compare the true limiting variance to the semiparametric efficiency bound for a general dimension of the covariate vector. We show that the post-matching DiD estimators do not attain the bound when the number of matches is fixed, but approach it as the number of neighbors diverges under some conditions.

We illustrate our results with simulations and with a reanalysis of the National Supported Work (NSW) Demonstration. Using the experimental benchmark from the randomized evaluation following \citet{Lalonde_1986_AER}, we find that the post-matching 2WFE estimator yields estimates that are much closer to the experimental estimates than the unmatched 2WFE, and accounting for the matching step increases the standard errors. 

Finally, while our main results focus on nearest-neighbor matching with replacement, we discuss how they extend to matching on discrete covariates, nearest-neighbor matching without replacement and propensity score matching.

Our paper contributes to a large and growing literature on DiD and 2WFE methods in staggered designs, that is, designs where treated units exhibit variation in their treatment timing \citep[see][for surveys]{Chaise-Dhaut_2022_EJ,Roth-etal_2023_JoE}. In particular, studies have found that in staggered designs, 2WFE estimators generally recover linear combinations of ATTs with weights that are hard to interpret and possibly negative when the treatment effects are heterogeneous \citep{Athey-Imbens_2021_JoE,Goodman-Bacon_2021_JoE,Chaise-Dhaut_2020_AER,Imai-Kim_2021_PA,Sun-Abraham_2021_JoE,Borusyak-Jaravel-Spiess_2024_ReStud}. These studies focus on unmatched 2WFE estimators, mainly under the unconditional version of the parallel trends assumption. We complement this literature by showing that these issues remain for post-matching 2WFE estimators when the parallel trends assumption holds conditionally. Furthermore, we show that pooling different treatment cohorts in the matching step creates an additional bias. Specifically, our Theorem \ref{thm:staggered} decomposes the probability limit of the post-matching 2WFE estimator into a linear combination of cohort-specific ATTs and three types of bias: one due to the heterogeneity of treatment effects over time, one due to the heterogeneity of treatment effects across cohorts and a bias due to differences in covariate distributions across different treated cohorts. Because the third bias, which is specific to our setting, is caused by differences in covariate distributions and not by treatment effect heterogeneity, this bias remains even when the treatment effect is constant.

Another closely related strand of the literature is the one analyzing DiD and 2WFE under a conditional parallel trends assumption \citep{Abadie_2005_ReStud,SantAnna-Zhao_2020_JoE,Callaway-SantAnna_2021_JoE}. These studies focus on inverse-probability weighting and doubly-robust estimators. We build on their framework to analyze alternative estimators that control for covariates employing fully non-parametric nearest-neighbor matching methods. 

Our paper also contributes to the literature on cross-sectional matching methods \citep{Heckman-Ichimura-Todd_1998_ReStud,Abadie-Imbens_2006_ECMA,Abadie-Imbens_2008_ECMA,Abadie-Imbens_2011_JBES,Abadie-Imbens_2012_JASA,Abadie-Imbens_2016_ECMA,Abadie-Spiess_2022_JASA} by generalizing the results to 2WFE estimators with panel data. Specifically, our Propositions \ref{prop:weighted_reg_full} and \ref{prop:weighted_reg} show that while the post-matching 2WFE that pools all treated cohorts is a complicated linear combination of 2x2 DiD estimators, the estimator that compares one treated cohort to the never-treated can be recast as a cross-sectional matching estimator using a simple transformation of the outcome of interest, and thus existing matching methods can be used to characterize the probability limit and asymptotic distribution of this estimator. In addition, our closed-form characterization of the asymptotic variance based on the results by \citet{Chen-Han_2024_wp} allows us to compare it to the semiparametric efficiency bound for a general dimension of the covariate vector, a result previously available only for the scalar case. In line with existing results for the average treatment effect (ATE) in cross-sectional settings \citep{Abadie-Imbens_2006_ECMA,Lin-Ding-Han_2021_ECMA}, we find that the asymptotic variance is larger than the semiparametric efficiency bound, but approaches it under some conditions as the number of neighbors diverges. 

Finally, among the papers at the intersection of matching and panel data methods, \citet{Heckman-Ichimura-Todd_1997_ReStud} propose a DiD-matching estimator in a two-period setting and derive its limiting distribution when matching is conducted using kernel and local linear regression methods. We complement their findings by considering nearest-neighbor matching with multiple periods and a staggered design, a setting that is very common in empirical work. \citet{Imai-Kim-Wang_2023_AJPS} propose matching methods for panel data, matching units based on their treatment history. They also show that their estimators can be written as weighted 2WFE regressions, but their inference is conducted conditionally on the matching weights and does not account for the variability introduced by the matching step.

The remainder of the paper is organized as follows. Section \ref{sec:setup} describes the setup, notation and provides identification conditions for the parameters of interest. Section \ref{sec:estimation} characterizes the estimators of interest and their probability limits. Section \ref{sec:asy_distr} studies the asymptotic distribution of the estimators, provides a closed-form formula for their limiting variance, discusses efficiency and the inconsistency of the naive cluster-robust variance estimator that ignores the matching step. Section \ref{sec:simul} contains simulations studies and Section \ref{sec:emp_app} illustrates our results using data from the NSW program. Section \ref{sec:extensions} discusses extensions of our results to other types of matching, and Section \ref{sec:conclusion} provides concluding remarks.


\section{Setup and Identification}\label{sec:setup}

Consider panel data following $n$ units $i=1,\ldots,n$ over $T$ time periods $t=1,2,\ldots,T$. Let $t_i^*$ be a random variable with support $\mathcal{S}=\{s\in\{2,\ldots,T,\infty\}:\P[t_i^*=s]>0\}$ that indicates the period in which unit $i$ receives the treatment for the first time. The values of $t_i^*$ identify different treatment cohorts. We assume that there is at least one pure pre-treatment period (and possibly more) in which no unit is treated, $\P[t_i^*=1]=0$, and denote the never-treated units by $t_i^*=\infty$. We denote the population proportion of each treatment cohort by $p_s=\P[t_i^*=s]$. We assume that the treatment is an absorbing state, so that once unit $i$ is treated, it remains treated through period $T$. Let $D_{it}=\1(t\ge t_i^*)$ be the treatment indicator in each period. 

For $t'>t$, let $\mathbf{d}_{t:t'}=(d_t,d_{t+1},\ldots,d_{t'})$ be a vector of treatments between periods $t$ and $t'$, where $d_{t}\in\{0,1\}$ and $d_{t+1}\ge d_t$ for all $t$, and let $\mathbf{0}_{t:t'}$ and $\mathbf{1}_{t:t'}$ be vectors with all their entries being equal to zero or one, respectively. We use the shorthand notation $\mathbf{d}_{1:t}=\mathbf{d}_t$. Without further assumptions, the potential outcome for each unit in each time period may be a function of the whole treatment path, $Y_{it}(\mathbf{d}_T)$. We introduce the following assumption, standard in the DiD literature, which ensures that potential outcomes do not depend on future treatments.

\begin{assumption}[No treatment anticipation]\label{assu:no_antic}
For all $t$, $Y_{it}(\mathbf{d}_T)=Y_{it}(\mathbf{d}_t)$.
\end{assumption}

We use the shorthand notation $Y_{it}(\mathbf{0}_t)=Y_{it}(0)$ to denote the potential outcome of a unit not receiving treatment up to period $t$ and, for any $t\ge t_i^*$, we write $Y_{it}(t_i^*)=Y_{it}(\mathbf{0}_{t_i^*-1},\mathbf{1}_{t^*_i:t})$, which is the potential outcome in period $t$ for a unit that enters treatment in period $t_i^*$. The unit-level treatment effect is defined as $\tau_{it}=Y_{it}(t_i^*)-Y_{it}(0)$. The main parameters of interest are the cohort-specific ATTs, $\E[\tau_{it}|t_i^*=s]$.

Let $X_i$ be a time-invariant vector of covariates of dimension $q\ge 1$, measured at a pure pre-treatment period $t<\min\{t_i^*\}$. To justify the matching procedure, we assume that the parallel trends assumption holds conditional on covariates in the following way.

\begin{assumption}[Conditional parallel trends]\label{assu:cpt}
For all $t$,
\[\E[Y_{it}(0)-Y_{it-1}(0)|t_i^*,X_i]=\E[Y_{it}(0)-Y_{it-1}(0)|X_i] \quad (a.s.).\]
\end{assumption}
This assumption states that for each value of $X_i$, all treatment cohorts experience the same outcome trend in the absence of treatment.


Finally, let $e_s(X_i)=\P[t_i^*=s|X_i]$ be the cohort-specific propensity score, that is, the probability of entering treatment in period $s$ given covariates $X_i$. We impose the following conditions which guarantees a non-zero mass of never treated and common support.

\begin{assumption}[Overlap]\label{assu:overlap}
$\P[t_i^*=\infty]>0$ and for all $s\in\mathcal{S}$, $e_s(X_i)<1-\eta$ (a.s.) for some $\eta>0$.
\end{assumption}

The following identification result is standard in the literature, and we include it here for reference.

\begin{lemma}\label{lemma:identif}
Under Assumptions \ref{assu:no_antic}, \ref{assu:cpt} and \ref{assu:overlap}, for any $t'<s \le t<s'$, 
\[\E[\tau_{it}|t_i^*=s]=\E[Y_{it}-Y_{it'}|t_i^*=s]-\E\left\{\left.\E[Y_{it}-Y_{it'}|t_i^*=s',X_i]\right\vert t_i^*=s\right\}.\]
\end{lemma}


\section{Post-matching 2WFE estimation}\label{sec:estimation}

Consider a sample $(Y_{i1},\ldots,Y_{iT},t_i^*,X_i')_{i=1}^n$. Let $N_s=\sum_i \1(t_i^*=s)$ be the number of units in treatment cohort $t_i^*=s<\infty$ and $N_0=\sum_i\1(t_i^*=\infty)$ be the number of never-treated units in the sample, with corresponding sample proportions $\hat{p}_s=N_s/n$ and $\hat{p}_\infty=N_0/n$. We assume observations are drawn independently from the same distribution.

\begin{assumption}[Random sampling]\label{assu:sampling}
$(Y_{i1},\ldots,Y_{iT},t_i^*,X_i')_{i=1}^n$ is a sample of independent and identically distributed observations from a population $(Y_1,\ldots,Y_T,t^*,X')$.
\end{assumption}

We study the following two-step estimation strategy commonly used in empirical practice. In the first step, each treated unit is matched with replacement to $M\ge1$ units from the pool of never-treated units based on their covariates $X_i$. 
In the second step, treatment effects are estimated using a 2WFE regression with weights defined by the matching procedure, as described in detail below.

To analyze the matching step, we rely on the setup from \citet{Abadie-Imbens_2006_ECMA}. As in their paper, we consider the case in which all the covariates are continuous, and discuss matching on discrete covariates in Section \ref{sec:extensions}. Let $\norm{x}=\sqrt{x'x}$. For each unit $i$, we define $j_m(i,s')$ as the index $j\in\{1,\ldots,n\}$ of units in comparison cohort $s'\ne t_i^*$ such that:
\[\sum_{l:t_l^*=s'}\1\left\{\norm{X_l-X_i}\le \norm{X_j-X_i}\right\}=m,\]
that is, the index of the $m$-th closest unit to $i$ in covariate values among the comparison cohort $s'$. Let $\mathcal{J}_M(i,s')=\{j_1(i,s'),j_2(i,s'),\ldots,j_M(i,s')\}$ be the set of indices for the $M$ nearest neighbors in cohort $s'$ for unit $i$. Finally, let:
\[K_M(i,s)=\sum_{l:t_l^*=s}\1\{i\in\mathcal{J}_M(l,t_i^*)\}\]
be the number of times unit $i$ is used as a match for units in cohort $s$. We also let:
\[K_M(i)=\sum_{s\ne t_i^*}K_M(i,s)\]
the total number of times unit $i$ is used as a match for units in any other cohort.

Once the matching is conducted, treatment effects are estimated using a 2WFE regression on the matched sample that includes the treated units and their nearest neighbors. We consider the following general weighted 2WFE estimator. Let $w_i$ be (possibly random) unit-level weights, collected in a vector $w=(w_1,\ldots,w_n)$. For each time period $t=1,\ldots,T$, let $\lambda_t\in\{0,1\}$ be a nonrandom indicator of whether the period is included in the analysis, and collect these indicators in a vector $\lambda=(\lambda_1,\ldots,\lambda_T)$ where $\sum_t\lambda_t\ge 2$ so there are at least two periods. The weighted 2WFE estimator is defined as the estimator of $\tau$ from the 2WFE regression:\footnote{We assume that the weights $(w,\lambda)$ are chosen in such a way that this estimator is well defined, which requires avoiding perfect collinearity between the regressor $D_{it}$ and the unit and time effects. This rules out obvious cases where, for example, no included cohort experiences a treatment change between the included periods, or all included cohorts experience the same treatment change.}
\begin{align}\label{eq:2wfe_reg}
Y_{it}=\alpha_i+\delta_t+\tau D_{it}+u_{it}
\end{align}
using weights $w_i\lambda_t$, and can be written as: 
\begin{align}\label{eq:2wfe}
\hat\tau(w,\lambda)&=\frac{\sum_i\sum_tw_i\lambda_tY_{it}(D_{it}-\bar{D}_i-\tilde{D}_t+\bar{D})}{\sum_i\sum_tw_i\lambda_tD_{it}(D_{it}-\bar{D}_i-\tilde{D}_t+\bar{D})}
\end{align}
where
\[\bar{D}_i=\frac{\sum_t\lambda_tD_{it}}{\sum_t\lambda_t},\quad \tilde{D}_t=\frac{\sum_iw_iD_{it}}{\sum_iw_i},\quad \bar{D}=\frac{\sum_i\sum_tw_i\lambda_tD_{it}}{\sum_iw_i\sum_t\lambda_t}.\]
Lemma \ref{lemma_app:2wfe} in the appendix provides a general representation of this estimator for an arbitrary set of weights. 

In what follows, we focus on the empirical approach described above, which assigns a weight equal to one to all treated units, pooling all treatment cohorts, and reweight the never-treated units by the number of times they are used as a match, scaled by the number of neighbors $M$. We refer to this estimator as the full-sample matched 2WFE estimator. The following result characterizes this estimator.

\begin{proposition}\label{prop:weighted_reg_full}
Let
\begin{align*}
w_i^\mathsf{pool}=\sum_{s\ne\infty}\1(t_i^*=s)+\1(t_i^*=\infty)\frac{K_M(i)}{M}
\end{align*}
and $w^\mathsf{pool}=(w^\mathsf{pool}_1,\ldots,w^\mathsf{pool}_n)$. With these weights, the weighted 2WFE estimator \eqref{eq:2wfe} can be written as:
\begin{align*}
\hat\tau(w^\mathsf{pool},\lambda)&=\frac{\hat\tau_\mathsf{num}^\mathsf{pool}(\lambda)}{\hat\tau_\mathsf{den}^\mathsf{pool}(\lambda)}
\end{align*}
where $\Lambda_s=\sum_{t\ge s}\lambda_t/\sum_t\lambda_t$,
\begin{align*}
\hat\tau_\mathsf{num}^\mathsf{pool}(\lambda)&=\sum_{s<\infty}\hat{p}_s\left\{\sum_{t\ge s}\sum_{t'<s}\lambda_t\lambda_{t'}\hat\tau^s_\infty(t,t')+\sum_{s'\ne s,\infty}\frac{\hat{p}_{s'}}{1-\hat{p}_\infty}\sum_{t\ge s}\sum_{t'<s}\lambda_t\lambda_{t'}\hat\tau^s_{s'}(t,t')\right\}\\
\hat\tau_\mathsf{den}^\mathsf{pool}(\lambda)&=\sum_{s<\infty}\hat{p}_s\Lambda_s\left\{(1-\Lambda_s)+\frac{\sum_{s'<s}\hat{p}_{s'}(\Lambda_{s'}-\Lambda_s)}{1-\hat{p}_\infty}\right\}+\sum_{s<s'<\infty}\frac{\sum_{s<\infty}\hat{p}_s\hat{p}_{s'}}{1-\hat{p}_\infty}(1-\Lambda_s)(\Lambda_s-\Lambda_{s'})\\
&\times\left(\sum_t\lambda_t\right)^2\\
\hat\tau^s_{\infty}(t,t')&=\frac{1}{N_s}\sum_i \left(\1(t_i^*=s)-\1(t_i^*=\infty)\frac{K_M(i,s)}{M}\right)\left(Y_{it}-Y_{it'}\right)\\
\hat\tau^s_{s'}(t,t')&=\frac{1}{N_s}\sum_i \1(t_i^*=s)\left(Y_{it}-Y_{it'}\right)-\frac{1}{N_{s'}}\sum_i\1(t_i^*=s')\left(Y_{it}-Y_{it'}\right),\quad s'\ne\infty.
\end{align*}
\end{proposition}

Proposition \ref{prop:weighted_reg_full} shows that the full-sample matched 2WFE estimator can be decomposed as a linear combination of all possible 2x2 estimators, comparing each cohort $s$ to the never-treated cohort in different post- and pre-treatment periods, $\hat\tau^s_{s'}(t,t')$, and comparing each treated cohort $s$ to other eventually treated cohorts, $\hat\tau^s_{s'}(t,t')$ for $s'\ne\infty$. As pointed out in the literature on unmatched 2WFE estimators  \citep{Chaise-Dhaut_2020_AER,Athey-Imbens_2021_JoE,Goodman-Bacon_2021_JoE,Imai-Kim_2021_PA,Sun-Abraham_2021_JoE,Borusyak-Jaravel-Spiess_2024_ReStud}, the second set of comparisons is problematic because it involves some comparison groups that are already treated, often known as ``forbidden comparisons''. Importantly, while the 2x2 estimators using the never-treated, $\hat\tau^s_\infty(t,t')$, use the matching weights, the 2x2 estimators comparing different eventually treated cohorts, $\hat\tau^s_{s'}(t,t')$ for $s'\ne\infty$, are unweighted DiD comparisons. These unweighted comparisons generally bias the estimator, as we show next. To characterize the probability limit of the estimator, we introduce the following regularity conditions. 

\begin{assumption}[Regularity conditions]\label{assu:reg}$ $
\begin{enumerate}
\item The conditional densities of $X_i|t_i^*=\infty$ and $X_i|t_i^*=s$ for all $s\in\mathcal{S}\setminus \{\infty\}$, $f_0(x)$ and $f_s(x)$ respectively, are continuous and have convex and compact support, and $f_0(x)$ is bounded away from zero.
\item For all $s$, $t$ and any treatment path $\mathbf{d}_t$, $\E[Y_{it}(\mathbf{d}_t)|t_i^*=s,X_i=x]$ is Lipschitz-continuous.
\item For all $s$, $t$ and any treatment path $\mathbf{d}_t$, $\V[Y_{it}(\mathbf{d}_t)|t_i^*=s,X_i=x]$ is Lipschitz-continuous and bounded away from zero.
\item For all $s$, $t$ and any treatment path $\mathbf{d}_t$, $\E[Y_{it}(\mathbf{d}_t)^4|t_i^*=s,X_i=x]$ is uniformly bounded.
\end{enumerate}
\end{assumption}

\begin{theorem}\label{thm:staggered}
Under Assumptions \ref{assu:no_antic}-\ref{assu:reg},
\begin{align*}
\hat\tau(w^\mathsf{pool},\lambda)\to_\P &\sum_{s<\infty}\phi_1(s,\lambda)\frac{\sum_{t\ge s}\lambda_t\E[\tau_{it}|t_i^*=s]}{\sum_{t\ge s}\lambda_t}+\sum_{s<\infty}\sum_{s<s'<\infty}\phi_2(s,s',\lambda)\frac{\sum_{s\le t<s'}\lambda_t\E[\tau_{it}|t_i^*=s]}{\sum_{s\le t<s'}\lambda_t}\\
+&B^\mathsf{het}_\mathsf{cohort}-B^\mathsf{het}_\mathsf{time}+B_\mathsf{pool}
\end{align*}
where $B^\mathsf{het}_\mathsf{cohort}$, $B^\mathsf{het}_\mathsf{time}$ and $B_\mathsf{pool}$ are bias terms defined as:
\begin{align*}
B^\mathsf{het}_\mathsf{cohort}&=\sum_{s<\infty}\sum_{s'\ne s,\infty}\eta_1(s,s',\lambda)\frac{\sum\limits_{t\ge \max\{s,s'\}}\lambda_t\left(\E[\tau_{it}|t_i^*=s]-\E[\tau_{it}|t_i^*=s']\right)}{\sum\limits_{t\ge \max\{s,s'\}}\lambda_t}\\
B^\mathsf{het}_\mathsf{time}&=\sum_{s<\infty}\sum_{s'<s}\eta_2(s,s',\lambda)\frac{\sum_{t\ge s}\sum_{s'\le t'<s}\lambda_t\lambda_{t'}\E[\tau_{it}-\tau_{it'}|t_i^*=s']}{\sum_{t\ge s}\sum_{s'\le t'<s}\lambda_t\lambda_{t'}}\\
B_\mathsf{pool}&=\sum_{s<\infty}\sum_{s'\ne s,\infty}\eta_3(s,s',\lambda)\frac{\sum_{t\ge s}\sum_{t'<s}\lambda_t\lambda_{t'}B_{s'}^s(t,t')}{\sum_{t\ge s}\sum_{t'<s}\lambda_t\lambda_{t'}}\\
B_{s'}^s(t,t')&=\E[Y_{it}(0)-Y_{it'}(0)|t_i^*=s]-\E[Y_{it}(0)-Y_{it'}(0)|t_i^*=s']
\end{align*}
and where, letting $\Lambda_s=\sum_{t\ge s}\lambda_t/\sum_t\lambda_t$, the weights are:
\begin{align*}
\phi_1(s,\lambda)&=\frac{p_s\Lambda_s\left[(1-\Lambda_s)+\sum_{s'<s}p_{s'}(\Lambda_{s'}-\Lambda_s)/(1-p_\infty)\right]}{\sum_{s<\infty}\left\{p_s\Lambda_s\left[(1-\Lambda_s)+\frac{\sum_{s'<s}p_{s'}(\Lambda_{s'}-\Lambda_s)}{1-p_\infty}\right]+\sum_{s<s'<\infty}\frac{p_sp_{s'}}{1-p_\infty}(1-\Lambda_s)(\Lambda_s-\Lambda_{s'})\right\}}\\
\phi_2(s,s',\lambda)&=\frac{p_sp_{s'}(1-\Lambda_s)(\Lambda_s-\Lambda_{s'})/(1-p_\infty)}{\sum_{s<\infty}\left\{p_s\Lambda_s\left[(1-\Lambda_s)+\frac{\sum_{s'<s}p_{s'}(\Lambda_{s'}-\Lambda_s)}{1-p_\infty}\right]+\sum_{s<s'<\infty}\frac{p_sp_{s'}}{1-p_\infty}(1-\Lambda_s)(\Lambda_s-\Lambda_{s'})\right\}}\\
\eta_1(s,s',\lambda)&=\frac{p_sp_{s'}\Lambda_{\max\{s,s'\}}(1-\Lambda_{\min\{s,s'\}})/(1-p_\infty)}{\sum_{s<\infty}\left\{p_s\Lambda_s\left[(1-\Lambda_s)+\frac{\sum_{s'<s}p_{s'}(\Lambda_{s'}-\Lambda_s)}{1-p_\infty}\right]+\sum_{s<s'<\infty}\frac{p_sp_{s'}}{1-p_\infty}(1-\Lambda_s)(\Lambda_s-\Lambda_{s'})\right\}}\\
\eta_2(s,s',\lambda)&=\frac{p_sp_{s'}\Lambda_s(\Lambda_{s'}-\Lambda_s)/(1-p_\infty)}{\sum_{s<\infty}\left\{p_s\Lambda_s\left[(1-\Lambda_s)+\frac{\sum_{s'<s}p_{s'}(\Lambda_{s'}-\Lambda_s)}{1-p_\infty}\right]+\sum_{s<s'<\infty}\frac{p_sp_{s'}}{1-p_\infty}(1-\Lambda_s)(\Lambda_s-\Lambda_{s'})\right\}}\\
\eta_3(s,s',\lambda)&=\frac{p_sp_{s'}\Lambda_s(1-\Lambda_s)/(1-p_\infty)}{\sum_{s<\infty}\left\{p_s\Lambda_s\left[(1-\Lambda_s)+\frac{\sum_{s'<s}p_{s'}(\Lambda_{s'}-\Lambda_s)}{1-p_\infty}\right]+\sum_{s<s'<\infty}\frac{p_sp_{s'}}{1-p_\infty}(1-\Lambda_s)(\Lambda_s-\Lambda_{s'})\right\}}.
\end{align*}
\end{theorem}

Theorem \ref{thm:staggered} shows that in a staggered design, the post-matching 2WFE estimator does not generally recover a causally interpretable parameter. The probability limit of this estimator can be decomposed into four components. The first component, in the first line of the expression for $\hat\tau(w^\mathsf{pool},\lambda)$ above, is a weighted average of ATTs across cohorts and over the included time periods, where the weights are non-negative and sum up to one, $\sum_{s<\infty}\phi_1(s,\lambda)+\sum_{s<\infty}\sum_{s<s'<\infty}\phi_2(s,s',\lambda)=1$. Although this is a proper weighted average, the weights may be highly unequal across ATTs, and this term is generally hard to interpret.

In addition to this weighted average of ATTs, the full-sample matched 2WFE estimator contains three bias terms. The first bias, $B^\mathsf{het}_\mathsf{cohort}$, involves differences in ATTs in a given time period between different cohorts, $\E[\tau_{it}|t_i^*=s]-\E[\tau_{it}|t_i^*=s']$. These terms are non-zero whenever treatment effects are heterogeneous across cohorts. Notice that because this term consists of differences in ATTs, some ATTs enter negatively. This is an illustration of the negative weights issue pointed out in the literature on unmatched DiD with staggered adoption.

The second bias, $B^\mathsf{het}_\mathsf{time}$, involves the change in the ATTs for a given cohort over time, $\E[\tau_{it}-\tau_{it'}|t_i^*=s']$. This term appears because the 2WFE estimator uses already-treated units as comparison units, resulting in some ATTs entering with a negative sign in the probability limit of the estimator. Notice that the bias terms $B^\mathsf{het}_\mathsf{cohort}$ and $B^\mathsf{het}_\mathsf{time}$ have been discussed in existing literature analyzing unmatched DiD estimators. Theorem \ref{thm:staggered} shows that these issues persist when using matched 2WFE estimators.

The third bias, $B_\mathsf{pool}$, on the other hand, is specific to the post-matching setting when the matching is conducted as explained above. This ``pooling bias'' involves differences in untreated outcome trends across eventually-treated cohorts, $B_{s'}^s(t,t')$. To understand the intuition behind this bias, notice that these terms can be rewritten as:
\[B_{s'}^s(t,t')=\E\left\{\E\left[Y_{it}(0)-Y_{it'}(0)|t_i^*=s,X_i\right]\left(\frac{f_s(X_i)-f_{s'}(X_i)}{f(X_i)}\right)\right\}\]
where $f(x)$ is the unconditional density of $X_i$ and $f_s(x)$ and $f_{s'}(x)$ are the conditional densities of $X_i$ for cohorts $s$ and $s'$, respectively. Thus, this bias is driven by the difference in covariate distributions across eventually-treated cohorts. As shown in Proposition \ref{prop:weighted_reg_full}, the full-sample matched 2WFE estimator involves comparisons between different treated cohorts. Although each treated unit is matched in the first step to a never-treated unit, ensuring that the distribution of covariates for the treated is identical to the reweighted distribution of the never-treated, the different treated cohorts are not matched to each other, and thus their covariate distributions will differ in general. Because the parallel trends assumption only holds conditionally on $X_i$, different treated cohorts may have different unconditional trends in their untreated outcomes, which creates this bias. In other words, when the parallel trends assumption holds conditionally, any unmatched comparison between eventually treated cohorts is invalid, even those involving not-yet-treated units as comparisons.

When the treatment effect is constant both across units and over time $\tau_{it}=\tau$, $\E[\tau_{it}|t_i^*=s]-\E[\tau_{it}|t_i^*=s']=\E[\tau_{it}-\tau_{it'}|t_i^*=s']=0$, so the biases due to treatment effect heterogeneity equal zero, $B^\mathsf{het}_\mathsf{time}=B^\mathsf{het}_\mathsf{cohort}=0$. However, the bias due to the difference in covariate distributions,  $B_\mathsf{pool}$, remains regardless of whether treatment effects are homogeneous or heterogeneous. This bias equals zero when either the parallel trends assumption holds unconditionally, $\E\left[Y_{it}(0)-Y_{it'}(0)|t_i^*=s,X_i\right]=\E\left[Y_{it}(0)-Y_{it'}(0)|t_i^*=s\right]$ for all $s$, so matching is unnecessary, or when the distribution of covariates across cohorts are the same, $f_s(x)=f_{s'}(x)$, which is generally not true even when the treatment effect is constant.

Thus, the commonly used full-sample matched 2WFE estimator requires stringent homogeneity conditions on treatment effects and covariate distributions to recover causally interpretable parameters when there is variation in treatment timing. To avoid this problem, a staggered design can be split into multiple pairwise designs, as suggested by \citet{Callaway-SantAnna_2021_JoE,Sun-Abraham_2021_JoE,Steigerwald-VazquezBare-Meier_2021_JAERE}, among others, in unmatched panel designs. We now consider an estimator that assigns weight equal to one to units in a specific cohort $t_i^*=s$, weights equal to $K_M(i,s)/M$ to never-treated units, and weight equal to zero to all other cohorts. We refer to these estimators as pairwise matched DiD estimators. The following proposition characterizes this estimator. 

\begin{proposition}\label{prop:weighted_reg}
For a given $s\in\mathcal{S}\setminus\{\infty\}$, let:
\begin{align*}
w_i^s=\1(t_i^*=s)+\1(t_i^*=\infty)\frac{K_M(i,s)}{M},
\end{align*}
with $w^s=(w_1^s,\ldots,w_n^s)$. With these weights, the weighted 2WFE estimator \eqref{eq:2wfe} can be written as:
\begin{align}\label{eq:tauhat}
\hat\tau(w^s,\lambda)&=\frac{1}{N_s}\sum_i \left(\1(t_i^*=s)-\1(t_i^*=\infty)\frac{K_M(i,s)}{M}\right)\left(\bar{Y}_i^{\mathsf{post},s}(\lambda)-\bar{Y}_i^{\mathsf{pre},s}(\lambda)\right)
\end{align}
where 
\[\bar{Y}_i^{\mathsf{post},s}(\lambda)=\frac{\sum_{t\ge s}\lambda_tY_{it}}{\sum_{t\ge s}\lambda_t},\quad \bar{Y}_i^{\mathsf{pre},s}(\lambda)=\frac{\sum_{t<s}\lambda_tY_{it}}{\sum_{t<s}\lambda_t}\]
are the average outcome for each unit in the included post- and pre-treatment periods, respectively.
\end{proposition}

Proposition \ref{prop:weighted_reg} shows that the pairwise matched DiD estimator can be recast as cross-sectional matching estimator like the ones studied by \citet{Abadie-Imbens_2006_ECMA}, where the outcome is the difference in average post- and pre-treatment periods. As an illustration, for a given $s$, when $\lambda_t^*=\1(t\in\{\ell',\ell\})$ for a pair of periods such that $\ell'<s\le \ell$,
\begin{align*}
\hat\tau(w^s,\lambda^*)&=\frac{1}{N_s}\sum_i \left(\1(t_i^*=s)-\1(t_i^*=\infty)\frac{K_M(i,s)}{M}\right)\left(Y_{i\ell}-Y_{i\ell'}\right)
\end{align*}
which is a 2x2 (i.e. two-group-two-period) matched DiD estimator.

\begin{remark}[Normalized weights]\label{rmk:weights}
Because weighted OLS automatically normalizes the weights to sum to one, and since the sum of the number of times each untreated unit is used as a match, $\sum_i\1(t_i^*=\infty)K_M(i,s)$, equals the total number of matches, $N_sM$, it follows that Proposition \ref{prop:weighted_reg} also holds for normalized weights:
\[\tilde{w}_i^s=\frac{\1(t_i^*=s)}{N_s}+\1(t_i^*=\infty)\frac{K_M(i,s)}{N_sM}\]
which replicate those in \citet{Abadie-Imbens_2006_ECMA}.
\end{remark}

The pairwise matched DiD estimator is consistent for an average of the ATTs for cohort $s$ over the included post-treatment periods $t\ge s$, as we show below.

\begin{theorem}\label{thm:consistency}
Under Assumptions \ref{assu:no_antic}-\ref{assu:reg},
\[\hat\tau(w^s,\lambda)\to_\P \frac{\sum_{t\ge s}\lambda_t \E[\tau_{it}|t_i^*=s]}{\sum_{t\ge s}\lambda_t}.\]
\end{theorem}
As an illustration, if $\lambda_t=\1(t\in\{\ell',\ell\})$ with $\ell'<s\le \ell$, we have that $\hat\tau(w^s,\lambda)\to_\P \E[\tau_{i\ell}|t_i^*=s]$ which is the ATT for cohort $s$ in period $\ell$. On the other hand, when all periods are included, so that $\lambda_t=1$ for all $t$, $\hat\tau(w^s,\lambda)\to_\P\sum_{t\ge s}\E[\tau_{it}|t_i^*=s]/(T+1-s)$ which is a simple average of ATTs over periods $t\ge s$.

\begin{remark}[Not-yet-treated]
This result, and all subsequent results for the pairwise matched DiD estimator, immediately generalizes to the case where units in cohort $s$ are matched to units in some other eventually-treated cohort $s'\ne s$ instead of the never-treated, as long as cohort $s'$ is untreated in all periods for which $\lambda_t=1$. This corresponds to the case where the not-yet-treated are used as comparison units, as discussed in \citet{Callaway-SantAnna_2021_JoE} and \citet{Sun-Abraham_2021_JoE}.
\end{remark}

These results demonstrate that the inconsistency of the full-sample matched 2WFE estimator can be avoided by comparing each treated cohort to the never treated, which consistently estimates an average of ATTs over time. When conducting inference based on the pairwise matched DiD estimators, the variance needs to account for the variability introduced by the matching step, and failing to account for this variability generally results in invalid standard error estimators. The next section studies the asymptotic distribution of the pairwise matched DiD estimators, proposes valid standard error estimators and characterizes the bias of the naive variance estimators that ignore the matching step.


\section{Asymptotic Distribution and Inference}\label{sec:asy_distr}

In this section, we study the asymptotic distribution and valid inference procedures for the pairwise estimators \eqref{eq:tauhat}. In what follows, for a given $s\in\mathcal{S}$, let $\Delta \bar{Y}_i^s(\lambda)=\bar{Y}^{\mathsf{post},s}_i(\lambda)-\bar{Y}^{\mathsf{pre},s}_i(\lambda)$, $\mu_\lambda(s,x)=\E[\Delta \bar{Y}_i^s(\lambda)|t_i^*=s,X_i=x]$ and $\sigma^2_\lambda(s,x)=\V[\Delta \bar{Y}_i^s(\lambda)|t_i^*=s,X_i=x]$ where we use the notation $\mu_\lambda(0,x)$ and $\sigma^2_\lambda(0,x)$ when $s=\infty$. Also let:
\[\bar{\tau}_i^{\mathsf{post},s}(\lambda)=\frac{\sum_{t\ge s}\lambda_t\tau_{it}}{\sum_{t\ge s}\lambda_t}\]
be the average unit-level treatment effect over the included post-treatment periods $t\ge s$ with $s<\infty$. The following result characterizes the asymptotic distribution of the pairwise matched DiD estimator.
\begin{theorem}\label{thm:asynorm}
Under Assumptions \ref{assu:no_antic}-\ref{assu:reg}, for $s\in\mathcal{S}\setminus\{\infty\}$,
\begin{align*}
\sqrt{n}\left(\hat\tau(w^s,\lambda)-\E[\bar{\tau}_i^{\mathsf{post},s}(\lambda)|t_i^*=s]-B_\lambda(s,M)\right)\to_\mathcal{D}\mathcal{N}(0,V(s,\lambda))
\end{align*}
where
\begin{align*}
V(s,\lambda)&= \frac{1}{p_s}\E\left[\left.\left(\E[\bar{\tau}_i^{\mathsf{post},s}(\lambda)|t_i^*=s,X_i]-\E[\bar{\tau}_i^{\mathsf{post},s}(\lambda)|t_i^*=s]\right)^2\right\vert t_i^*=s\right]+\frac{1}{p_s}\E[\sigma_\lambda^2(s,X_i)|t_i^*=s]\\
&+\frac{1}{p_s^2M^2}\E\left[\left(M+\alpha(M,q)\frac{e_s(X_i)}{e_\infty(X_i)}\right)e_s(X_i)\sigma_\lambda^2(0,X_i)\right],
\end{align*}
$\alpha(M,q)$ is a distribution-free, nonrandom function defined in \citet{Chen-Han_2024_wp} and 
\begin{align*}
B_\lambda(s,M)&=\frac{1}{N_s}\sum_i\left(\1(t_i^*=s)-\1(t_i^*=\infty)\frac{K_M(i,s)}{M}\right)\mu_\lambda(0,X_i).
\end{align*}
\end{theorem}

Theorem \ref{thm:asynorm} shows that the pairwise matched DiD estimator is asymptotically normal after appropriate centering and scaling, which means that valid inference can be conducted using standard procedures. In general, the estimator needs to be bias-corrected, as initially pointed out by \citet{Abadie-Imbens_2006_ECMA}. We refer to this bias $B_\lambda(s,M)$ as the ``matching bias''. The matching bias $B_\lambda(s,M)$ is a finite-sample bias due to the fact that, when covariates are continuous, it is not possible to find perfect matches for treated units in any particular sample. This type of finite sample bias is typical in nonparametric estimators. It can be shown that under the assumptions in the theorem, $B_\lambda(s,M)=O_\P(n^{-1/q})$ where $q$ is the dimension of $X_i$. This implies that $B_\lambda(s,M)=o_\P(1)$ so the matching bias does not appear in the probability limit of the estimator, but it needs to be accounted for in the asymptotic distribution because it is multiplied by the convergence rate. 

In some cases, the bias correction is not needed. For example, when matching on a single covariate, $\sqrt{n}B_\lambda(s,M)=o_\P(1)$. \citet{Abadie-Imbens_2006_ECMA} also provide an alternative sampling scheme in which the bias is asymptotically negligible when the never-treated group is ``sufficiently larger'' than the treated group; see Section \ref{sec:wo_repl} for further discussion. Finally, when matching on discrete covariates, matches are exact, so the bias is zero, as discussed in Section \ref{sec:discrete}.

For these reasons, the nature of the matching bias is very different from that of the pooling bias of the full-sample matched 2WFE estimator characterized in Theorem \ref{thm:staggered},  $B_\mathsf{pool}$. Specifically, the pooling bias is a large-sample bias that appears in the probability limit of the estimator, and does not decrease with the sample size.

Because the pairwise estimator is asymptotically normal, valid inference can be conducted using standard methods after estimating the bias and the asymptotic variance that accounts for the matching step. Proposition \ref{prop:weighted_reg} shows that the pairwise matched DiD estimator can be written as a cross-sectional estimator, and therefore the methods available in the literature for bias and variance estimation like the ones in \citet{Abadie-Imbens_2006_ECMA,Abadie-Imbens_2011_JBES} can be applied in our setting after transforming the data appropriately.


\subsection{Efficiency}

The analytic formula for the asymptotic variance we derive in Theorem \ref{thm:asynorm} allows us to study the efficiency of the pairwise matched DiD estimator for a general dimension of the covariate vector. Under Assumptions \ref{assu:cpt}, \ref{assu:overlap} and \ref{assu:sampling}, \citet{SantAnna-Zhao_2020_JoE} derive the semiparametric efficiency bound for the ATT with panel data under the conditional parallel trends assumption. When comparing cohort $s$ against the never-treated, the bound is given by:
\begin{align}\label{eq:speb}
\begin{split}
V_\mathsf{SEB}(\lambda)&=\E\left[\frac{\sigma^2_\lambda(s,X_i)e_s(X_i)}{p_s^2}+\frac{(\E[\bar{\tau}_i^{\mathsf{post},s}(\lambda)|t_i^*=s,X_i]-\E[\bar{\tau}_i^{\mathsf{post},s}(\lambda)|t_i^*=s])^2e_s(X_i)}{p_s^2}\right]\\
&+\E\left[\frac{e_s(X_i)^2\sigma^2_\lambda(0,X_i)}{p_s^2e_\infty(X_i)}\right].
\end{split}
\end{align}
Then we have the following result.
\begin{proposition}\label{prop:seb}
\begin{align*}
V(s,\lambda)-V_\mathsf{SEB}(\lambda)&=\frac{1}{M}\E\left[\frac{e_s(X_i)\sigma^2_\lambda(0,X_i)}{p_s^2}\right]+\left(\frac{\alpha(M,q)}{M^2}-1\right)\E\left[\frac{e_s(X_i^2)\sigma^2_\lambda(0,X_i)}{e_\infty(X_i)p_s^2}\right].
\end{align*}
\end{proposition}

Proposition \ref{prop:seb} shows that the matched DiD estimator does not attain the semiparametric efficiency bound, and that the difference between the asymptotic variance and the bound depends on the number of nearest neighbors $M$ and on the dimension of the covariates $q$ through the function $\alpha(M,q)$. As an illustration, with a single covariate $q=1$, $\alpha(M,1)=M(2M+1)/2$ \citep{Chen-Han_2024_wp} and thus:
\begin{align*}
V(s,\lambda)-V_\mathsf{SEB}&=\frac{1}{M}\E\left[\frac{e_s(X_i)\sigma^2_\lambda(0,X_i)}{p_s^2}\right]+\frac{1}{2M}\E\left[\frac{e_s(X_i^2)\sigma^2_\lambda(0,X_i)}{e_\infty(X_i)p_s^2}\right].
\end{align*}
In this case it is clear that the difference between the variance and the bound vanishes as $M\to\infty$, in line with the results in \citet{Abadie-Imbens_2006_ECMA} and \citet{Lin-Ding-Han_2021_ECMA} for the ATE with cross-sectional data under unconfoundedness. Because the closed form of $\alpha(M,q)$ has not been characterized for a general $M$ and $q$, it is hard to extend this statement for the general case. However, the tabulations in \citet{Chen-Han_2024_wp} show that for $M\in\{2,3,\ldots10\}$ and $d\in\{2,3,\ldots,10\}$, $\alpha(M,q)/M^2-1<0.15$, and that $\alpha(M,q)/M^2-1$ is positive and decreasing in $M$ for each of these values of $M$ and $q$. 



\subsection{Inference Ignoring the Matching Step}

Because the pairwise matched DiD estimator is constructed in two steps, inference needs to account for the first-step estimation in the second-step asymptotic variance, as done in Theorem \ref{thm:asynorm}. In practice, however, researchers often rely on the cluster-robust variance estimator from the second-stage 2WFE regression, failing to account for the variability introduced by the matching step, which results in inconsistent variance estimators. With cross-sectional data under unconfoundedness, \citet{Abadie-Imbens_2016_ECMA} compare the asymptotic variance of the oracle propensity-score matching ATT estimator with the variance of the ATT estimator using the estimated propensity score, and show that the latter can be smaller or larger depending on the data generating process. On the other hand, \citet{Abadie-Spiess_2022_JASA} consider post-matching regression estimators using cross-sectional data and matching without replacement and show that standard errors that fail to account for the matching step are inconsistent when the regression of interest is misspecified. In this section, we derive the probability limit of the cluster-robust 2WFE variance estimator that ignores the matching step, which we refer to as the naive cluster-robust variance estimator.

By Proposition \ref{prop:weighted_reg}, the cluster-robust 2WFE estimator of the variance of $\hat\tau(w^s,\lambda)$ can be written as:
\begin{align*}
\hat{V}_\mathsf{ncr}&=\frac{1}{N_s^2}\sum_i\1(t_i^*=s)\hat\varepsilon_i^2+\frac{1}{N_s^2}\sum_i\1(t_i^*=\infty)\frac{K_M(i,s)^2}{M^2}\hat\varepsilon_i^2
\end{align*}
where:
\begin{align*}
\hat\varepsilon_i=\Delta \bar{Y}_i^s(\lambda)&-\left(\frac{1}{N_s}\sum_j\1(t_j^*=s)\Delta \bar{Y}_j^s(\lambda)\right)\1(t_i^*=s)\\
&-\left(\frac{1}{N_s}\sum_j\1(t_j^*=\infty)\frac{K_M(j,s)}{M}\Delta \bar{Y}_j^s(\lambda)\right)\1(t_i^*=\infty)
\end{align*}
is the regression residual and the dependence on $\lambda$ is left implicit to reduce notation. The following result characterizes the probability limit of this estimator.

\begin{proposition}\label{prop:cluster_se}
Under Assumptions \ref{assu:no_antic}-\ref{assu:reg},
\begin{align*}
n\hat{V}_\mathsf{ncr}&\to_\P V(s,\lambda)\\
&+\frac{1}{p_s}\V[\E[\Delta\bar{Y}_i^s(0,\lambda)|t_i^*=s,X_i]|t_i^*=s]\\
&+\frac{2}{p_s}\cov(\E[\bar{\tau}_i^{\mathsf{post},s}(\lambda)|t_i^*=s,X_i],\E[\Delta\bar{Y}_i^s(0,\lambda)|t_i^*=s,X_i]|t_i^*=s)\\
&+\frac{1}{p_s^2M^2}\E\left[\left(M+\alpha(M,q)\frac{e_s(X_i)}{e_\infty(X_i)}\right)e_s(X_i)(\E[\Delta\bar{Y}_i^s(0,\lambda)|t_i^*=s,X_i]-\E[\Delta\bar{Y}_i^s(0,\lambda)|t_i^*=s])^2\right].
\end{align*}
where
\[\Delta\bar{Y}_i^s(0,\lambda)=\frac{\sum_{t\ge s}\lambda_tY_{it}(0)}{\sum_{t\ge s}\lambda_t}-\frac{\sum_{t<s}\lambda_tY_{it}(0)}{\sum_{t<s}\lambda_t}.\]
\end{proposition}

Proposition \ref{prop:cluster_se} shows that if $\E[\Delta\bar{Y}_i^s(0,\lambda)|t_i^*=s,X_i]=\E[\Delta\bar{Y}_i^s(0,\lambda)|t_i^*=s]$, so that covariates do not affect the average counterfactual trend, the naive cluster-robust variance estimator that ignores the matching step is consistent for the true variance. This case, however, only occurs when the matching step is unnecessary, because counterfactual trends are parallel unconditionally. 
More generally, the naive cluster-robust variance estimator is inconsistent for the true asymptotic variance, with an asymptotic bias given by:
\begin{align*}
&\frac{1}{p_s}\V[\E[\Delta\bar{Y}_i^s(0,\lambda)|t_i^*=s,X_i]|t_i^*=s]\\
&+\frac{2}{p_s}\cov(\E[\bar{\tau}_i^{\mathsf{post},s}(\lambda)|t_i^*=s,X_i],\E[\Delta\bar{Y}_i^s(0,\lambda)|t_i^*=s,X_i]|t_i^*=s)\\
&+\frac{1}{p_s^2M^2}\E\left[\left(M+\alpha(M,q)\frac{e_s(X_i)}{e_\infty(X_i)}\right)e_s(X_i)(\E[\Delta\bar{Y}_i^s(0,\lambda)|t_i^*=s,X_i]-\E[\Delta\bar{Y}_i^s(0,\lambda)|t_i^*=s])^2\right].
\end{align*}
The direction and magnitude of this bias depend on the data generating process. The first and third and terms are always positive, whereas the covariance can be positive or negative. For example, if the conditional ATT, $\E[\tau_{it}|t_i^*=s,X_i]$, is constant over $X_i$, the covariance is zero and thus the naive variance estimator is upward biased. In such cases, ignoring the matching step results in overly conservative standard errors. On the other hand, when the covariance between the conditional ATT and the conditional trend of the untreated outcome is negative and large enough, the bias may be negative, resulting in standard errors that underestimate the true variability of the estimator. We further illustrate these cases with simulations in the next section.




\section{Simulations}\label{sec:simul}

We now present two simulation studies to illustrate the problems with post-matching 2WFE estimators analyzed in previous sections. In the first study, we consider a staggered treatment adoption setting to illustrate the asymptotic bias. In the second study, we show how failing to account for the matching step when conducting inference results in invalid standard errors, considering cases where the unadjusted standard errors are too large and too small.


\subsection{Bias in Staggered Designs}

Our first setup consists of panel data with a total of $n=1,000$ units and four time periods, $t=1,2,3,4$. Units belong to one of three cohorts, $t_i^*\in\{2,3,\infty\}$, so the ``early adopters'' enter treatment in period $t=2$ and the ``late adopters'' enter treatment in period $t=3$. Units are assigned to cohorts based on a multinomial rule:
\[\P[t_i^*=\infty|X_i]=\frac{1}{1+\exp(X_i)+\exp(2X_i)},\quad \P[t_i^*=s|X_i]=\frac{\exp((s-1)X_i)}{1+\exp(X_i)+\exp(2X_i)},\quad s=2,3\]
where $ X_i$ is a scalar covariate with $X_i\sim \mathrm{Uniform}(-1,2)$. The untreated potential outcome is defined as:
\[Y_{it}(0)=\alpha_i+(t-1)+5X_i(t-1)+u_{it},\quad \alpha_i\sim\mathcal{N}(0,1),\quad u_{it}\sim \mathcal{N}(0,1)\]
with $(\alpha_i,u_{it},X_i)$ mutually independent. This implies that $\E[Y_{it}(0)-Y_{it-1}(0)|t_i^*=s,X_i]=1+5X_i$ for all $s$, so the conditional parallel trends assumption holds (the unconditional parallel trends assumption does not because $\E[X_i|t_i^*=s]$ differs across $s$). To focus on the bias due to the matching procedure and abstract from the other biases characterized in Theorem \ref{thm:staggered}, we assume that the treatment effect is constant, $\tau=5$. The model for the treated potential outcome is:
\[Y_{it}(1)=5 + \alpha_i+(t-1)+5X_i(t-1)+v_{it},\quad v_{it}\sim \mathcal{N}(0,1).\]
We estimate $\tau$ by matching each treated unit to a never-treated unit in the first step and running Equation \eqref{eq:2wfe_reg} in the reweighted data. The results are displayed in the first column of Table \ref{tab:simul_results}. The results show that the bias of the full sample matched 2WFE is about 0.7 or 15\% of the true ATT. In addition, the naive cluster-robust standard error severely overestimates the true variability.


\subsection{Post-Matching Inference}

Our second simulation exercise illustrates how failing to account for the matching step leads to invalid inference. We consider panel data with a total of $n=1,000$ units and four time periods, $t=1,2,3,4$. Units belong to one of two cohorts, $t_i^*\in\{3,\infty\}$, so there is a treated cohort that enters treatment in period 3, and a never-treated cohort. Units are assigned to treatment based on a logistic model:
\[\P[t_i^*=3|X_i]=\frac{\exp(X_i)}{1+\exp(X_i)}\]
where $X_i$ is a scalar covariate with $X_i\sim\mathrm{Uniform}[-1/2,1/2]$. We consider two designs. The first design has a constant treatment effect equal to 5 and the potential outcomes are given by:
\begin{align*}
Y_{it}(0)&=\alpha_i+(t-1)+5X_i(t-1)+u_{it}\\
Y_{it}(1)&=5+\alpha_i + (t-1)+5X_i(t-1)+v_{it}
\end{align*}
where as before $u_{it}\sim \mathcal{N}(0,1)$, $v_{it}\sim \mathcal{N}(0,1)$ and $(\alpha_i,u_{it},X_i)$ mutually independent. Because in this setup the treatment effect is constant, Proposition \ref{prop:cluster_se} shows that the naive cluster-robust standard error is biased upwards.

In the second design, the potential outcomes are given by:
\begin{align*}
Y_{it}(0)&=\alpha_i+(t-1)-2X_i(t-1)+u_{it}\\
Y_{it}(1)&=5+\alpha_i + (t-1) +5X_i(t-1)+v_{it}
\end{align*}
which implies that the conditional ATT is $\E[Y_{it}(1)-Y_{it}(0)|t_i^*=3,X_i]=7X_i(t-1)$ which changes across $X_i$ and over time. On the other hand, $\E[\Delta Y_{it}(0)|X_i,t_i^*=s]=1-2X_i$ and $\cov(\E[Y_{it}(1)-Y_{it}(0)|t_i^*=3,X_i],\E[\Delta Y_{it}(0)|X_i,t_i^*=s]|t_i^*=s)=-14(t-1)\V[X_i]$ which in this case results in downward-biased standard errors.

The results from these simulations are shown Table \ref{tab:simul_results}. Column (2) reports the result for the constant treatment effect scenario. As shown in Proposition \ref{prop:cluster_se}, when the treatment effect is constant, the naive variance estimator overestimates the true variability. In this case, the Monte Carlo (MC) standard error is 0.086, while the average of the naive standard error across simulations is 0.257, about three times larger. This upward bias results in overcoverage of confidence intervals and loss of power. On the other hand, the standard error estimator that accounts for the variability of the matching procedure is 0.085, almost identical to the MC standard error, and reaches correct coverage.

Column (3) reports the results for the case with a heterogeneous treatment effect and a negative covariance between the conditional ATT and the conditional outcome trend under no treatment, the third term in Proposition \ref{prop:cluster_se}. In this case, the negative covariance term yields a downward-biased standard error of 0.209, about 13\% smaller than the MC standard error (0.239) and the naive CI undercovers the true parameter. As before, the standard error that accounts for the matching procedure gives is very close to the MC standard error and reaches correct coverage.

\begin{table}[htbp]
\caption{Simulations Results}\label{tab:simul_results}
\begin{center}
\begin{tabular}{lrrr}
\hline\hline
\multicolumn{1}{l}{}&\multicolumn{1}{c}{(1)}&\multicolumn{1}{c}{(2)}&\multicolumn{1}{c}{(3)}\tabularnewline
\hline
Bias&0.702&-0.000&-0.002\tabularnewline
MC se&0.195&0.086&0.239\tabularnewline
Naive CR se&1.274&0.257&0.209\tabularnewline
Coverage (naive)&100.0&100.0&91.2\tabularnewline
CR se&-&0.085&0.239\tabularnewline
Coverage&-&94.3&95.0\tabularnewline
\hline
\end{tabular}

\end{center}

\footnotesize{\textbf{Notes}: ``Bias'' is the average difference between the estimates and the true parameter across simulations. ``MC se'' is the Monte Carlo standard standard deviation of the estimates. ``Naive CR se'' is the average naive cluster-robust standard error that ignores the matching step, and ``Coverage (naive)'' is the coverage of the corresponding 95\% confidence interval. ``CR se'' is the cluster-robust standard error that accounts for the matching step, and ``Coverage'' is the coverage of the corresponding 95\% confidence interval. Results from 5,000 simulations.}
\end{table}


\section{Empirical Application}\label{sec:emp_app}

We now illustrate our results using data from the National Supported Work (NSW) Demonstration. Using the NSW experimental data as a benchmark, \citet{Lalonde_1986_AER} compared different non-experimental methods to determine whether they can replicate the experimental findings. This study sparked an extensive debate about the credibility of non-experimental methods \citep[see][for a recent review]{Imbens-Xu_2025_JEL}.

Following \cite{Lalonde_1986_AER}, we use data from the Current Population Survey (CPS) to construct a comparison group for the experimental treatment group using nearest-neighbor matching. The outcome of interest is earnings, which is measured in 1974, 1975 (before treatment) and 1978 (after treatment). Our list of covariates includes indicators of age at baseline, indicators of years of education and indicators of being married, Black and Hispanic.

We compare the following specifications using earnings data from 1975 and 1978, which constitutes a two-period panel: (1) a difference in mean earnings between treated and controls in the experimental sample (``Experimental''), (2) an unmatched 2WFE regression (``2WFE''), (3) a naive matched 2WFE regression where standard errors do not account for the matching step (``Naive matched 2WFE''), (4) a matched 2WFE regression calculating valid standard errors (``Matched 2WFE''), (5) a matched 2WFE regression with valid standard errors and bias correction using a linear specification for the covariates (``Matched 2WFE - BC'') and (6) a 2WFE regression on the matched sample, that is, on the sample excluding any untreated units that are not matched to any treated unit, but without accounting for the number of times each untreated is used as a match (``2WFE in matched sample''). All standard errors are clustered at the individual level.

The results are shown in Column (1) of Table \ref{tab:results}. The experimental estimate of the ATT is \$886.3 with a p-value of 0.069. We use this value as a benchmark. The unmatched 2WFE regression yields an estimate of about \$1,714, almost twice as big as the experimental benchmark, with a p-value close to zero. The weighted 2WFE that uses matching in the first step gives an estimate of around \$930 (p-value 0.117), much closer to the experimental benchmark. As shown in Proposition \ref{prop:cluster_se}, the standard error estimator in this specification is inconsistent because it does not account for the matching step. When adjusting for the matching first step, the standard error increases from \$593 to \$635. The bias-corrected matched 2WFE specification reports a slightly lower estimate of \$809.21 with a very similar standard error. Finally, the 2WFE regression dropping unmatched units reveals an estimate of about \$418, approximately half the experimental benchmark.

Column (2) of Table \ref{tab:results} repeats the same analysis for a placebo outcome. In the experimental sample, this corresponds to a difference in mean earnings in 1975 (before the treatment) between treated and controls. For all other specifications, the placebo test uses data on earnings in 1974 and 1975, and can be interpreted as a pre-trends test. Because all the data in these specifications is pre-treatment, all estimated effects should be close to zero and insignificant. While all the estimates are insignificant at the usual levels, the one from the unweighted 2WFE and the one from the regression that ignores repeated matches show the largest magnitudes and lowest p-values.

\begin{table}[htbp]
\caption{Estimated Effects of the NSW Job Training Program}\label{tab:results}
\begin{center}
\begin{tabular}{lcc}
\hline\hline
\multicolumn{1}{l}{}&\multicolumn{1}{c}{(1) ATT estimate}&\multicolumn{1}{c}{(2) Placebo test}\tabularnewline
\hline
{\bfseries Experiment}&&\tabularnewline
~~Coefficient&886.30&39.42\tabularnewline
~~Std. err.&488.14&379.01\tabularnewline
~~p-value&0.069&0.917\tabularnewline
~~95\% CI&[-70.45 ; 1,843.06]&[-703.44 ; 782.27]\tabularnewline
\hline
{\bfseries 2WFE}&&\tabularnewline
~~Coefficient&1,714.40&-138.90\tabularnewline
~~Std. err.&485.93&179.79\tabularnewline
~~p-value&\textless  0.001&0.440\tabularnewline
~~95\% CI&[761.98 ; 2,666.81]&[-491.30 ; 213.49]\tabularnewline
\hline
{\bfseries Naive matched 2WFE}&&\tabularnewline
~~Coefficient&929.82&44.20\tabularnewline
~~Std. err.&593.26&309.65\tabularnewline
~~p-value&0.117&0.886\tabularnewline
~~95\% CI&[-232.97 ; 2,092.62]&[-562.71 ; 651.11]\tabularnewline
\hline
{\bfseries Matched 2WFE}&&\tabularnewline
~~Coefficient&929.82&44.20\tabularnewline
~~Std. err.&635.38&339.08\tabularnewline
~~p-value&0.143&0.896\tabularnewline
~~95\% CI&[-315.52 ; 2,175.17]&[-620.39 ; 708.79]\tabularnewline
\hline
{\bfseries Matched 2WFE - BC}&&\tabularnewline
~~Coefficient&809.21&219.10\tabularnewline
~~Std. err.&636.85&342.10\tabularnewline
~~p-value&0.204&0.522\tabularnewline
~~95\% CI&[-439.02 ; 2,057.43]&[-451.42 ; 889.61]\tabularnewline
\hline
{\bfseries 2WFE in matched sample}&&\tabularnewline
~~Coefficient&418.10&-306.91\tabularnewline
~~Std. err.&521.98&215.93\tabularnewline
~~p-value&0.423&0.155\tabularnewline
~~95\% CI&[-604.99 ; 1,441.19]&[-730.13 ; 116.31]\tabularnewline
\hline
\end{tabular}

\end{center}

\footnotesize{\textbf{Notes}: the table show the estimates from (1) a difference in mean earnings between treated and controls in the experimental sample (``Experimental''), (2) an unmatched 2WFE regression (``2WFE''), (3) a naive matched 2WFE regression where standard errors do not account for the matching step (``Naive matched 2WFE''), (4) a matched 2WFE regression calculating valid standard errors (``Matched 2WFE''), (5) a matched 2WFE regression with valid standard errors and bias correction using a linear specification for the covariates (``Matched 2WFE - BC'') and (6) a 2WFE regression on the matched sample, that is, on the sample excluding any untreated units that are not matched to any treated unit, but without accounting for the number of times each untreated is used as a match (``2WFE in matched sample''). All standard errors are clustered at the individual level.}
\end{table}


\section{Extensions to Other Types of Matching}\label{sec:extensions}

\subsection{Matching on Discrete Covariates}\label{sec:discrete}

When the vector $X_i$ is a low-dimensional vector of discrete covariates, the matching can be done exactly, which eliminates the matching bias. Without loss of generality, suppose that $X_i$ is a scalar taking values in a finite set $\mathcal{X}$. For such cases, a consistent and asymptotically normal estimator is given by:
\begin{align*}
\hat{\tau}_\mathsf{disc}(\lambda)&=\frac{1}{N_s}\sum_i\left(\1(t_i^*=s)-\1(t_i^*=\infty)\frac{\hat{e}_s(X_i)}{\hat{e}_\infty(X_i)}\right)\Delta \bar{Y}_i^s(\lambda)
\end{align*}
where
\[\hat{e}_s(X_i)=\hat{\P}[t_i^*=s|X_i=x]=\frac{\sum_i \1(t_i^*=s)\1(X_i=x)}{\sum_i\1(X_i=x)}.\]
This estimator that reweights never-treated units by the ratio of the estimated propensity scores is equivalent to the matching DiD estimator \eqref{eq:tauhat} that matches each treated unit to all never-treated units in the same covariate cell $X_i=x$. Because in this setting the number of units in each covariate cell increases as $n\to\infty$, this estimator can be seen as a matching estimator with a diverging number of matches, similar to the ATE estimator considered in a cross-sectional setting under unconfoundedness by \citet{Lin-Ding-Han_2021_ECMA}. This estimator attains the semiparametric efficiency bound \eqref{eq:speb}, as shown in the following result.

\begin{proposition}\label{prop:discrete}
Under Assumptions \ref{assu:no_antic}-\ref{assu:sampling}, if $E[Y_{it}(\mathbf{d}_t)^4|t_i^*=s,X_i=x]$ is bounded and $\V[Y_{it}(\mathbf{d}_t)|t_i^*=s,X_i=x]>0$ for all $\mathbf{d}_t$, $s$ and $x$,
\[\sqrt{n}(\hat{\tau}_\mathsf{disc}(\lambda)-\E[\bar{\tau}_i^{\mathsf{post},s}(\lambda)|t_i^*=s])\to_\mathcal{D}\mathcal{N}(0,V_\mathsf{SEB}(\lambda)).\]
\end{proposition}


\subsection{Matching Without Replacement}\label{sec:wo_repl}

Matching is sometimes carried out without replacement. In such cases, whenever an untreated unit is matched to a treated unit, that unit is removed from the pool of untreated. This implies that each treated unit is used at most once, $K_M(i)\in\{0,1\}$, and the size of the matched sample is $(M+1)N_s$. We now discuss how to adapt our results to the case of matching without replacement, building on the results by \citet{Abadie-Imbens_2012_JASA} and \citet{Abadie-Spiess_2022_JASA}. For simplicity we consider the case where the population consists of only two cohorts, $t_i^*\in\{s,\infty\}$. In this case, the post-matching 2WFE estimator using matching without replacement is:
\begin{align*}
\hat\tau_\mathsf{nr}(\lambda)&=\frac{1}{N_s}\sum_i \1(t_i^*=s)\left(\Delta Y^s_i(\lambda)-\frac{1}{M}\sum_{j\in\mathcal{J}_M(i,\infty)}\Delta Y^s_j(\lambda)\right).
\end{align*}
The following proposition summarizes the behavior of the ATT estimator under matching without replacement. 

\begin{proposition}\label{prop:wo_repl}
Suppose that the sample is obtained by drawing $N_s$ treated observations and $N_0$ untreated observations independently from the conditional distribution $Y_1,\ldots,Y_T,X|t^*=s$ and $Y_1,\ldots,Y_T,X|t^*=\infty$, respectively, and $N_s=O(N_0^{1/r})$ for some $r>q/2$. Then, under Assumptions \ref{assu:no_antic}, \ref{assu:cpt}, \ref{assu:overlap} and $\ref{assu:reg}$,
\begin{align*}
\sqrt{(M+1)N_s}(\hat{\tau}_\mathsf{nr}(\lambda)-\E[\bar{\tau}_i^{\mathsf{post},s}(\lambda)|t_i^*=s])\to_\mathcal{D}\mathcal{N}(0,V_\mathsf{nr}(\lambda))
\end{align*}
where:
\begin{align*}
V_\mathsf{nr}(\lambda)&=(1+M)\E\left[\left.\left(\E[\bar{\tau}_i^{\mathsf{post},s}(\lambda)|t_i^*=s,X_i]-\E[\bar{\tau}_i^{\mathsf{post},s}(\lambda)|t_i^*=s]\right)^2\right\vert t_i^*=s\right]\\
&+(1+M)\E\left[\left.\sigma^2_\lambda(s,X_i)+\frac{\sigma^2_\lambda(0,X_i)}{M}\right\vert t_i^*=s\right].
\end{align*}
\end{proposition}

The conditional sampling scheme in Proposition \ref{prop:wo_repl} and the requirement that $N_s=O(N_0^{1/r})$, which implies that the untreated group is ``sufficiently larger'' than the treated group, ensures that the bias of the matching estimator is $o_\P(n^{-1/2})$ \citep{Abadie-Imbens_2006_ECMA} and thus bias correction is not required for asymptotic normality. This is not specific to matching without replacement, and can also be applied to the results in the previous sections of the paper.


\subsection{Propensity Score Matching}

When the number of (continuous) covariates is large, matching directly on the vector of covariates can be computationally cumbersome. In such cases, researchers often resort to matching on the propensity score, which reduces the problem to matching on a single covariate. \citet{Rosenbaum-Rubin_1983_BIO} showed that the propensity score is a balancing score, and thus matching on the propensity score instead of the full vector of covariates is sufficient for eliminating confounding bias under a selection on observables assumption. In practice, the propensity score is unknown and needs to be estimated, which introduces additional variability that needs to be accounted for when conducting inference. \citet{Abadie-Imbens_2016_ECMA} derive the asymptotic distribution of the cross-sectional propensity score matching estimator. For the sake of brevity, we do not elaborate on the technical aspects of this case, but we note that the results in \citet{Abadie-Imbens_2016_ECMA} can be adapted to our setting to derive the asymptotic distribution of our pairwse matched DiD estimator when matching on the estimated propensity score.


\section{Conclusion}\label{sec:conclusion}

We analyze the commonly employed empirical strategy of estimating 2WFE regressions in a reweighted sample where treated units are matched to never-treated units in a first step. This approach is often used when the parallel trends assumption is believed to hold conditionally on covariates. By formally characterizing the post-matching 2WFE estimators and their asymptotic distribution, we highlight two problems with this strategy that generally invalidates its findings. First, when treated units exhibit variation in treatment timing, the post-matching 2WFE estimator that pools all treated units contains an asymptotic bias that remains even when the treatment effect is constant. The reason for this bias is that the 2WFE estimator implicitly involves comparisons between different treatment cohorts, using one of them as a comparison group. Because different treated cohorts are not matched to each other, they generally have different covariate distributions, and thus comparing their outcome evolutions over time is invalid when the parallel trends assumption holds conditionally, even when the comparison group is not yet treated.

Second, ignoring the matching step when conducting inference based on the post-matching 2WFE estimators results in inconsistent variance estimators, which can be biased upwards or downwards depending on the data generating process. Specifically, the naive variance estimator is upward-biased (conservative) when the treatment effect is constant, but can be downward-biased when there is sufficient treatment effect heterogeneity.

Then, we provide conditions under which the post-matching 2WFE estimator that compares each treated cohort to the never treated separately is consistent for an average of cohort-specific ATTs over time, and derive a closed-form representation of its asymptotic variance which can be used to conduct valid inference post-matching. We also show that this estimator does not generally attain the semiparametric efficiency bound, but it approaches it as the number of neighbors grows. We also illustrate our results using simulations and a reanalysis of the data from the NSW job training program. 

While different methods are available to incorporate covariates when the parallel trends assumption holds conditionally, matching-based methods have the advantage of being intuitively appealing, easy to implement, computationally stable and fully nonparametric. Our findings enable empirical researchers to exploit these methods and conduct valid inference on the parameters of interest. Moreover, because we show that post-matching 2WFE estimators can be recast as cross-sectional matching estimators, existing statistical software for matching estimators such as the \texttt{teffects nnmatch} package in Stata can be used after appropriately transforming the data.


\newpage

\bibliographystyle{econometrica}
\bibliography{post_matching_DiD_references}

\newpage

\appendix
\appendixpage

\renewcommand{\theapplemma}{\thesection\arabic{applemma}}

\renewcommand{\thetable}{A\arabic{table}}
\setcounter{table}{0}


\section{Survey of Empirical Literature}

Table \ref{tab:matching_ddd_literature} lists 22 papers published between 2018 and 2025 in top Economics and Political Science journals using some version of the post-matching 2WFE estimators analyzed in this paper. The journals we consider are, in alphabetical order: American Economic Journal: Applied Economics (AEJEP), American Economic Journal: Economic Policy (AEJEP), American Economic Review (AER), American Political Science Review (APSR) and Quarterly Journal of Economics (QJE).

\begin{sidewaystable}[htbp]
\centering
\caption{Survey of Empirical Literature}\label{tab:matching_ddd_literature}

\begin{tabular}{p{0.3 cm} p{4.2cm} p{9.5cm} p{1.3cm} p{2.0cm} p{5.0cm}}
\hline
&Author(s) & Title & Year & Journal & Matching type \\
\hline
1 & Powell, Seabury & Medical Care Spending and Labor Market Outcomes: Evidence from Workers’ Compensation Reforms  & 2018 & AER &
nearest neighbor matching \\
2 & Alfaro-urena, Manelici, Vasquez & The Effects of Joining Multinational Supply Chains: New Evidence from Firm to Firm Linkages  & 2022 & QJE &
nearest neighbor matching, propensity score matching \\
3 & Aneja and Xu & The Costs of Employment Segregation: Evidence from the Federal Government under Woodrow Wilson & 2022 & QJE &
coarsened Exact Matching \\
4 & Adams, Huttunen, Nix, Zhang & Violence Against Women at Work & 2024 & QJE &
nearest neighbour matching \\
5 & Adams, Huttunen, Nix, Zhang & The Dynamics of Abusive Relationships & 2024 & QJE &
nearest neighbour matching, propensity score matching \\
6 & Goldschmidt and Schmieder & The Rise of Domestic Outsourcing and the Evolution of the German Wage Structure & 2017 & QJE &
nearest neighbour matching, propensity score matching \\
7 & Cooper, Craig, Gaynor, Reenen & The Price Ain't Right? Hospital Prices and Health Spending on the Privately Insured & 2019 & QJE &
nearest neighbour matching, mahalanobis matching \\
8 & Aghamolla, Karaca-Mandic, Li, Thakor & Merchants of Death: The Effect of Credit Supply Shocks on Hosptial Outcomes & 2024 & AER &
propensity score matching \\
9 & Aneja, Luca, Reshef &   The Benefits of Revealing Race: Evidence from Minority-Owned Local Businesses & 2025 & AER &
coarsened exact matching \\
\hline
\end{tabular}

\end{sidewaystable}

\begin{sidewaystable}[!htbp]
\centering
\addtocounter{table}{-1}
\caption{Survey of Empirical Literature (continued)}

\begin{tabular}{p{0.3 cm} p{4.2cm} p{9.5cm} p{1.3cm} p{2.0cm} p{5.0cm}}
\hline
&Author(s) & Title & Year & Journal & Matching type \\
\hline
10 & Fenizia and Saggio & Organized Crime and Economic Growth: Evidence from Municipalities Infiltrated by the Mafia & 2024 & AER &
propensity score matching \\
11 & Sviatschi & Spreading Gangs: Exporting US Criminal Capital to El Salvador & 2022 & AER &
propensity score matching \\
12 & Schmieder, Wachter, Heining  & The Costs of Job Displacement over the Business Cycle
and Its Sources: Evidence from Germany & 2023 & AER &
propensity score matching \\
13 & Fetzer  & Did Austerity Cause Brexit? & 2019 & AER &
propensity score matching \\
14 & Muehlenbachs, Spiller, Timmins  & The Housing Market Impacts of Shale Gas Development & 2015 & AER &
nearest neighbour matching \\
15 & Fadlon, Nielsen  & Family Health Behaviors & 2019 & AER &
propensity score matching \\
16 & Hvide, Jones & University Innovation and the Professor’s Privilege & 2018 & AER &
propensity score matching \\
17 & Sigurdsson & Labor Supply Responses and Adjustment Frictions: A Tax-Free Year in Iceland & 2018 & AEJEP & coarsened exact matching \\
18 & Deryugina, MacKay, Reif & The Long-Run Dynamics of Electricity Demand: Evidence from Municipal Aggregation & 2020 & AEJAE & nearest neighbour matching \\
19 & Matsa, Miller & A Female Style in Corporate Leadership?
Evidence from Quotas & 2013 & AEJAE & nearest neighbour matching \\
20 & Ben-Manachem, Morris & Ticketing and Turnout: The Participatory Consequences of
Low-Level Police Contact & 2023 & APSR & nearest neighbour matching \\
21 & Incerti & How Firms, Bureaucrats, and Ministries Benefit from the Revolving
Door: Evidence from Japan & 2025 & APSR & nearest neighbour matching\\
22 & Rosenstiel & The Distributive Politics of Grants-in-Aid & 2025 & APSR & nearest neighbour matching\\
\hline
\end{tabular}

\end{sidewaystable}


\section{Auxiliary results}

\begin{applemma}\label{lemma_app:2wfe}
Let
\[\hat\tau(w,\lambda)=\frac{\sum_{i=1}^n\sum_{t=1}^Tw_iY_{it}(D_{it}-\bar{D}_i-\tilde{D}_t+\bar{D})}{\sum_{i=1}^n\sum_{t=1}^Tw_iD_{it}(D_{it}-\bar{D}_i-\tilde{D}_t+\bar{D})}\]
where $D_{it}=\1(t\ge t_i^*)$ and
\begin{align*}
\bar{D}_i=\frac{\sum_t\lambda_tD_{it}}{\sum_t\lambda_t},\quad \tilde{D}_t=\frac{\sum_iw_iD_{it}}{\sum_iw_i},\quad 
\bar{D}=\frac{\sum_i\sum_tw_i\lambda_tD_{it}}{\sum_iw_i\sum_t\lambda_t}.
\end{align*}
Then $\hat\tau(w,\lambda)=\hat\tau_\mathsf{num}(w,\lambda)/\hat\tau_\mathsf{den}(w,\lambda)$ where:
\begin{align*}
\hat\tau_\mathsf{num}(w,\lambda)&=\sum_{s<\infty}\sum_{s'\ne s}\sum_{t\ge s}\sum_{t'<s}\lambda_t\lambda_{t'}\hat{p}_s^w\hat{p}_{s'}^w\hat\tau_{s'}^s(t,t')\\
\hat\tau_\mathsf{den}(w,\lambda)&=\left(\sum_t\lambda_t\right)^2\sum_{s<\infty}\hat{p}_s^w\left\{\hat{p}_\infty^w\Lambda_s(1-\Lambda_s)+\sum_{s'<s}\hat{p}_{s'}^w\Lambda_s(\Lambda_{s'}-\Lambda_s)+\sum_{s<s'<\infty}\hat{p}_{s'}^w(1-\Lambda_s)(\Lambda_s-\Lambda_{s'})\right\}\\
\hat\tau_{s'}^s(t,t')&=\frac{\sum_iw_i\1(t_i^*=s)(Y_{it}-Y_{it'})}{\sum_iw_i\1(t_i^*=s)}-\frac{\sum_iw_i\1(t_i^*=s')(Y_{it}-Y_{it'})}{\sum_iw_i\1(t_i^*=s')}\\
\hat{p}_s^w&=\frac{\sum_iw_i\1(t_i^*=s)}{\sum_iw_i}.
\end{align*}
\end{applemma}


\begin{applemma}\label{lemma_app:conv}
Consider a subset of the data with two cohorts $s$ and $s'$, where each unit in cohort $s$ is matched to $M$ nearest neighbors from cohort $s'$. Let $(Z_i)_{i=1}^n$ be a sequence of iid random variables with $\E[Z_i|t_i^*=s,X_i=x]=\mu_Z(s,x)$ and $\V[Z_i|t_i^*=s,X_i=x]=\sigma^2_Z(s,x)$ with both functions being Lipschitz-continuous and with $\E[Z_i^4|t_i^*=s,X_i=x]$ uniformly bounded over $x$ for $s$ and $s'$. Under Assumption \ref{assu:overlap} and \ref{assu:reg}(1),
\[\frac{1}{N_s}\sum_i\1(t_i^*=s')\frac{K_M(i,s)}{M}Z_i\to_\P \E[\mu_Z(s',X_i)|t_i^*=s]\]
and
\[\frac{1}{N_s}\sum_i\1(t_i^*=s')\frac{K_M(i,s)^2}{M^2}Z_i^2\to_\P \frac{1}{M}\E\left[e_s(X_i)\frac{\E[Z_i^2|t_i^*=s',X_i]}{p_s}\right]+\frac{\alpha(M,q)}{M^2}\E\left[\frac{e_s(X_i)^2}{e_{s'}(X_i)}\frac{\E[Z_i^2|t_i^*=s',X_i]}{p_s}\right].\]
\end{applemma}


\section{Proofs of Main Results}

\subsection*{Proof of Lemma \ref{lemma:identif}}

\begin{align*}
\E[Y_{it}-Y_{it'}|t_i^*=s]&=\E[Y_{it}(t_i^*)-Y_{it'}(0)|t_i^*=s]\\
&=\E[\tau_{it}|t_i^*=s]+\E[Y_{it}(0)-Y_{it'}(0)|t_i^*=s]\\
&=\E[\tau_{it}|t_i^*=s]+\E\left\{\left.\E[Y_{it}(0)-Y_{it'}(0)|t_i^*=s,X_i]\right\vert t_i^*=s\right\}\\
&=\E[\tau_{it}|t_i^*=s]+\E\left\{\left.\E[Y_{it}(0)-Y_{it'}(0)|t_i^*=s',X_i]\right\vert t_i^*=s\right\}\\
&=\E[\tau_{it}|t_i^*=s]+\E\left\{\left.\E[Y_{it}-Y_{it'}|t_i^*=s',X_i]\right\vert t_i^*=s\right\}
\end{align*}
as required. $\square$

\subsection*{Proof of Proposition \ref{prop:weighted_reg_full}}
When $w_i^*=\sum_{s\ne\infty}\1(t_i^*=s)+\1(t_i^*=\infty)K_M(i)/M$, 
\begin{align*}
\sum_iw_i^*\1(t_i^*=s)&=N_s,\quad \sum_iw_i^*\1(t_i^*=\infty)=\frac{1}{M}\sum_i\1(t_i^*=\infty)K_M(i)=\sum_{s\ne\infty}N_s=n-N_0
\end{align*}
which implies that $\sum_iw_i^*=2(n-N_0)$, $\hat{p}_\infty^w=1/2$ and $\hat{p}_s^w=N_s/(2(n-N_0))=\hat{p}_s/(2(1-\hat{p}_\infty))$ for $s\ne\infty$, and the result follows by Lemma \ref{lemma_app:2wfe}. $\square$


\subsection*{Proof of Theorem \ref{thm:staggered}}

The numerator of the estimator can be decomposed as:
\begin{align*}
\hat\tau_\mathsf{num}^\mathsf{pool}(\lambda)&=\sum_{s<\infty}\sum_{t\ge s}\sum_{t'<s}\lambda_t\lambda_{t'}\hat{p}_s\hat\tau^s_\infty(t,t')\\
&+\sum_{s<\infty}\sum_{s<s'<\infty}\sum_{s\le t<s'}\sum_{t'<s}\lambda_t\lambda_{t'}\frac{\hat{p}_s\hat{p}_{s'}}{1-\hat{p}_\infty}\hat\tau^s_{s'}(t,t')\\
&+\sum_{s<\infty}\sum_{s'<s}\sum_{t\ge s}\sum_{t'<s'}\lambda_t\lambda_{t'}\frac{\hat{p}_s\hat{p}_{s'}}{1-\hat{p}_\infty}\hat\tau^s_{s'}(t,t')\\
&+\sum_{s<\infty}\sum_{s'<s}\sum_{t\ge s}\sum_{s'\le t'<s}\lambda_t\lambda_{t'}\frac{\hat{p}_s\hat{p}_{s'}}{1-\hat{p}_\infty}\hat\tau^s_{s'}(t,t')\\
&+\sum_{s<\infty}\sum_{s<s'<\infty}\sum_{t\ge s'}\sum_{t'<s}\lambda_t\lambda_{t'}\frac{\hat{p}_s\hat{p}_{s'}}{1-\hat{p}_\infty}\hat\tau^s_{s'}(t,t').
\end{align*}
Of these five set of comparisons, only the first two use comparison units are untreated in both periods. This information is summarized in the table below.

\begin{center}
\begin{tabular}{l|lll}
Comparison               & Cohort  & Status in $t'$ & Status in $t$ \\ \hline\hline
$t'<s\le t$, $s'=\infty$ & Never   & Untreated      & Untreated     \\
$t'<s\le t<s'<\infty$    & Later   & Untreated      & Untreated     \\
$t'<s<s'\le t$           & Later   & Untreated      & Treated       \\
$t'<s'<s\le t$           & Earlier & Untreated      & Treated       \\
$s'\le t'<s\le t$        & Earlier & Treated        & Treated       \\ \hline
\end{tabular}
\end{center}
Let
\begin{align*}
B_{s'}^s(t,t')&=\E[Y_{it}(0)-Y_{it'}(0)|t_i^*=s]-\E[Y_{it}(0)-Y_{it'}(0)|t_i^*=s'],
\end{align*}
By Lemma \ref{lemma_app:conv} and Lemma \ref{lemma:identif},
\begin{align*}
\hat\tau_{s'}^s(t,t')&\to_\P \E[\tau_{it}|t_i^*=s]\1(s'>t)+\left(\E[\tau_{it}|t_i^*=s]-\E[\tau_{it}-\tau_{it'}|t_i^*=s']\right)\1(s'\le t')\\
&+\left(\E[\tau_{it}|t_i^*=s]-\E[\tau_{it}|t_i^*=s']\right)\1(t'<s'\le t)\\
&+B_{s'}^s(t,t')\1(s'\ne\infty)
\end{align*}
and by the law of large numbers $\hat{p}_s\to_\P p_s$. Then,
\begin{align*}
\hat\tau_\mathsf{num}^\mathsf{pool}(\lambda)&\to_\P\sum_{s<\infty}p_s\sum_{t\ge s}\lambda_t\E[\tau_{it}|t_i^*=s]\left(\sum_{t'<s}\lambda_{t'}\right)\\
&+\sum_{s<\infty}\sum_{s<s'<\infty}\sum_{s\le t<s'}\sum_{t'<s}\lambda_t\lambda_{t'}\frac{p_sp_{s'}}{1-p_\infty}\left(\E[\tau_{it}|t_i^*=s]+B_{s'}^s(t,t')\right)\\
&+\sum_{s<\infty}\sum_{s'<s}\sum_{t\ge s}\sum_{t'<s'}\lambda_t\lambda_{t'}\frac{p_sp_{s'}}{1-p_\infty}\left(\E[\tau_{it}|t_i^*=s]-\E[\tau_{it}|t_i^*=s']+B_{s'}^s(t,t')\right)\\
&+\sum_{s<\infty}\sum_{s'<s}\sum_{t\ge s}\sum_{s'\le t'<s}\lambda_t\lambda_{t'}\frac{p_sp_{s'}}{1-p_\infty}\left(\E[\tau_{it}|t_i^*=s]-\E[\tau_{it}-\tau_{it'}|t_i^*=s']+B_{s'}^s(t,t')\right)\\
&+\sum_{s<\infty}\sum_{s<s'<\infty}\sum_{t\ge s'}\sum_{t'<s}\lambda_t\lambda_{t'}\frac{p_sp_{s'}}{1-p_\infty}\left(\E[\tau_{it}|t_i^*=s]-\E[\tau_{it}|t_i^*=s']+B_{s'}^s(t,t')\right).
\end{align*}
Rearranging,
\begin{align*}
\hat\tau_\mathsf{num}^\mathsf{pool}(\lambda)&\to_\P\sum_{s<\infty}p_s\Lambda_s(1-\Lambda_s)\frac{\sum_{t\ge s}\lambda_t\E[\tau_{it}|t_i^*=s]}{\sum_{t\ge s}\lambda_t}\left(\sum_t\lambda_t\right)^2\\
&+\sum_{s<\infty}\sum_{s'<s}\frac{p_sp_{s'}}{1-p_\infty}\Lambda_s(\Lambda_{s'}-\Lambda_s)\frac{\sum_{t\ge s}\lambda_t\E[\tau_{it}|t_i^*=s]}{\sum_{t\ge s}\lambda_t}\left(\sum_t\lambda_t\right)^2\\
&+\sum_{s<\infty}\sum_{s<s'<\infty}\frac{p_sp_{s'}}{1-p_\infty}(1-\Lambda_s)(\Lambda_s-\Lambda_{s'})\frac{\sum_{s\le t<s'}\lambda_t\E[\tau_{it}|t_i^*=s]}{\sum_{s\le t<s'}\lambda_t}\left(\sum_t\lambda_t\right)^2\\
&+\sum_{s<\infty}\sum_{s'\ne s,\infty}\frac{p_sp_{s'}}{1-p_\infty}\Lambda_{\max\{s,s'\}}(1-\Lambda_{\min\{s,s'\}})\frac{\sum\limits_{t\ge \max\{s,s'\}}\lambda_t\left(\E[\tau_{it}|t_i^*=s]-\E[\tau_{it}|t_i^*=s']\right)}{\sum\limits_{t\ge \max\{s,s'\}}\lambda_t}\left(\sum_t\lambda_t\right)^2\\
&-\sum_{s<\infty}\sum_{s'<s}\frac{p_sp_{s'}}{1-p_\infty}\Lambda_s(\Lambda_{s'}-\Lambda_s)\frac{\sum_{t\ge s}\sum_{s'\le t'<s}\lambda_t\lambda_{t'}\E[\tau_{it}-\tau_{it'}|t_i^*=s']}{\sum_{t\ge s}\sum_{s'\le t'<s}\lambda_t\lambda_{t'}}\left(\sum_t\lambda_t\right)^2\\
&+\sum_{s<\infty}\sum_{s'\ne s,\infty}\frac{p_sp_{s'}}{1-p_\infty}\Lambda_s(1-\Lambda_s)\frac{\sum_{t\ge s}\sum_{t'<s}\lambda_t\lambda_{t'}B_{s'}^s(t,t')}{\sum_{t\ge s}\sum_{t'<s}\lambda_t\lambda_{t'}}\left(\sum_t\lambda_t\right)^2.
\end{align*}
On the other hand, for the denominator,
\begin{align*}
\hat\tau_\mathsf{den}^\mathsf{pool}(\lambda)&\to\sum_{s<\infty}p_s\left\{\Lambda_s(1-\Lambda_s)+\sum_{s'<s}\frac{p_{s'}}{1-p_\infty}\Lambda_s(\Lambda_{s'}-\Lambda_s)+\sum_{s<s'<\infty}\frac{p_{s'}}{1-p_\infty}(1-\Lambda_s)(\Lambda_s-\Lambda_{s'})\right\}\\
&\times\left(\sum_t\lambda_t\right)^2
\end{align*}
and the result follows after simplifying the term $\left(\sum_t\lambda_t\right)^2$. $\square$

\subsection*{Proof of Proposition \ref{prop:weighted_reg}}

When $w_i^*=\1(t_i^*=s)+\1(t_i^*=\infty)K_M(i,s)/M$,
\begin{align*}
\sum_iw_i^*\1(t_i^*=s)&=N_s,\quad \sum_iw_i^*\1(t_i^*=\infty)=\frac{1}{M}\sum_i\1(t_i^*=\infty)K_M(i,s)=N_s
\end{align*}
using that $\sum_i\1(t_i^*=\infty)K_M(i,s)=N_sM$ because the number of times each untreated unit is used as a match equals the total number of matched units for the treated. This implies that $\sum_iw_i^*=2N_s$ and $\hat{p}_\infty^w=\hat{p}_s^w=1/2$ and the result follows by Lemma \ref{lemma_app:2wfe}. $\square$


\subsection*{Proof of Theorem \ref{thm:consistency}}

This result follows from Lemma \ref{lemma:identif} and Lemma \ref{lemma_app:conv} under conditional parallel trends. $\square$


\subsection*{Proof of Theorem \ref{thm:asynorm}}

Define $\Delta \bar{Y}_i^s(\lambda)=\bar{Y}^{\mathsf{post},s}_i-\bar{Y}^{\mathsf{pre},s}$, $\mu_\lambda(s,x)=\E[\Delta \bar{Y}_i^s(\lambda)|t_i^*=s,X_i=x]$ and $\sigma^2_\lambda(s,x)=\V[\Delta \bar{Y}_i^s(\lambda)|t_i^*=s,X_i=x]$ where we use the notation $\mu_\lambda(0,x)$ and $\sigma^2_\lambda(0,x)$ when $s=\infty$. Also let
\[\bar{\tau}_i^{\mathsf{post},s}(\lambda)=\frac{\sum_{t\ge s}\lambda_t\tau_{it}}{\sum_{t\ge s}\lambda_t}\]
By Assumption \ref{assu:cpt},
\[\E[\bar{\tau}_i^{\mathsf{post},s}(\lambda)|t_i^*=s,X_i]=\mu_\lambda(s,X_i)-\mu_\lambda(0,X_i).\]
By Proposition \ref{prop:weighted_reg}, the weighted estimator is:
\begin{align*}
\hat\tau(w^s,\lambda)&=\frac{1}{N_s}\sum_i \left(\1(t_i^*=s)-\1(t_i^*=\infty)\frac{K_M(i,s)}{M}\right)\Delta \bar{Y}_i^s(\lambda)
\end{align*}
This estimator can be rewritten as:
\begin{align*}
\hat\tau(w^s,\lambda)&=\frac{1}{N_s}\sum_i\1(t_i^*=s)(\mu_\lambda(s,X_i)-\mu_\lambda(0,X_i))\\
&+\frac{1}{N_s}\sum_i\left(\1(t_i^*=s)-\1(t_i^*=\infty)\frac{K_M(i,s)}{M}\right)(\Delta \bar{Y}_i^s(\lambda)-\mu_\lambda(t_i^*,X_i))\\
&+B_\lambda(s,M)
\end{align*}
where 
\begin{align*}
B_\lambda(s,M)&=\frac{1}{N_s}\sum_i\left(\1(t_i^*=s)-\1(t_i^*=\infty)\frac{K_M(i,s)}{M}\right)\mu_\lambda(0,X_i)\\
&=\frac{1}{N_s}\sum_i\1(t_i^*=s)\frac{1}{M}\sum_{m=1}^M\left(\mu_\lambda(0,X_i)-\mu_\lambda(0,X_{j_m(i,\infty)})\right).
\end{align*}
Then,
\begin{multline*}
\sqrt{n}\left(\hat\tau(w^s,\lambda)-\E[\bar{\tau}_i^{\mathsf{post},s}(\lambda)|t_i^*=s]-B_\lambda(s,M)\right)=\\
\frac{n}{N_s}\left\{\frac{1}{\sqrt{n}}\sum_i \1(t_i^*=s)\left(\mu_\lambda(s,X_i)-\mu_\lambda(0,X_i)-\E[\bar{\tau}_i^{\mathsf{post},s}(\lambda)|t_i^*=s]\right)\right\}\\
+\frac{n}{N_s}\left\{\frac{1}{\sqrt{n}}\sum_i\left(\1(t_i^*=s)-\1(t_i^*=\infty)\frac{K_M(i,s)}{M}\right)\varepsilon_i(t_i^*,X_i)\right\}
\end{multline*}
where $\varepsilon_i(t_i^*,X_i)=\Delta \bar{Y}_i^s(\lambda)-\mu_\lambda(t_i^*,X_i)$. Define:
\begin{align*}
Z_{n,i}&=\begin{cases}
\frac{1}{\sqrt{n}}\1(t_i^*=s)\left(\mu_\lambda(s,X_i)-\mu_\lambda(0,X_i)-\E[\bar{\tau}_i^{\mathsf{post},s}(\lambda)|t_i^*=s]\right) & \text{if }1\le i\le n\\
\frac{1}{\sqrt{n}}\left(\1(t_{i-n}^*=s)-\1(t_{i-n}^*=\infty)\frac{K_M(i-n,s)}{M}\right)\varepsilon_{i-n}(t_{i-n}^*,X_{i-n}) & \text{if } n+1\le i\le n
\end{cases}
\end{align*}
so that
\[\sqrt{n}\left(\hat\tau(w^s,\lambda)-\E[\bar{\tau}_i^{\mathsf{post},s}(\lambda)|t_i^*=s]-B_\lambda(s,M)\right)=\left(\frac{1}{p_s}+o_\P(1)\right)\sum_{i=1}^{2n}Z_{n,i}.\]
Define the $\sigma$-fields:
\begin{align*}
\mathcal{F}_{n,i}&=\begin{cases}
\sigma(\tstar,X_1,\ldots,X_i) &\text{if } 1\le i\le n\\
\sigma(\tstar,\X,\Delta \bar{Y}_1^s(\lambda),\ldots,\Delta \bar{Y}_{i-n}^s(\lambda)) &\text{if } n+1\le i\le 2n.
\end{cases}
\end{align*}
Then, the sequence
\[\left\{\left(\sum_{j=1}^i Z_{n,j},\mathcal{F}_{n,i}\right),1\le i\le 2n\right\}\]
is a martingale. Now, for $1\le i\le n$,
\begin{align*}
\E[Z_{n,i}^2|\mathcal{F}_{n,i}]&=\E\left[\left.\left(\frac{1}{\sqrt{n}}\1(t_i^*=s)\left(\mu_\lambda(s,X_i)-\mu_\lambda(0,X_i)-\E[\bar{\tau}_i^{\mathsf{post},s}(\lambda)|t_i^*=s]\right)\right)^2\right\vert\tstar,X_1,\ldots,X_{i-1}\right]\\
&=\frac{\1(t_i^*=s)}{n}\E\left[\left.\left(\mu_\lambda(s,X_i)-\mu_\lambda(0,X_i)-\E[\bar{\tau}_i^{\mathsf{post},s}(\lambda)|t_i^*=s]\right)^2\right\vert t_i^*=s\right]
\end{align*}
and for $n+1\le i\le 2n$,
\begin{align*}
\E[Z_{n,i}^2|\mathcal{F}_{n,i}]&=\E\left[\left.\left(\frac{1}{\sqrt{n}}\left(\1(t_{i-n}^*=s)-\1(t_{i-n}^*=s')\frac{K_M(i-n,s)}{M}\right)\varepsilon_{i-n}(t_{i-n}^*,X_{i-n})\right)^2\right\vert\mathcal{F}_{n,i-1}\right]\\
&=\frac{1}{n}\left(\1(t_{i-n}^*=s)-\1(t_{i-n}^*=s')\frac{K_M(i-n,s)}{M}\right)^2\E\left[\left.\varepsilon^2_{i-n}(t_{i-n}^*,X_{i-n})\right\vert t_{i-n}^*,X_{i-n}\right]\\
&=\frac{\1(t_{i-n}^*=s)}{n}\sigma_\lambda^2(s,X_{i-n})+\frac{\1(t_{i-n}^*=\infty)}{n}\cdot\frac{K_M(i-n,s)^2}{M^2}\sigma_\lambda^2(0,X_{i-n}).
\end{align*}
It follows that:
\begin{align*}
\sum_{i=1}^{2n}\E[Z_{n,i}^2|\mathcal{F}_{n,i}]&=\frac{N_s}{n}\E\left[\left.\left(\mu_\lambda(s,X_i)-\mu_\lambda(0,X_i)-\E[\bar{\tau}_i^{\mathsf{post},s}(\lambda)|t_i^*=s]\right)^2\right\vert t_i^*=s\right]\\
&+\frac{1}{n}\sum_{i=n+1}^{2n}\1(t_{i-n}^*=s)\sigma_\lambda^2(s,X_{i-n})+\frac{1}{n}\sum_{i=n+1}^{2n}\1(t_{i-n}^*=\infty)\frac{K_M(i-n,s)^2}{M^2}\sigma_\lambda^2(0,X_{i-n})\\
&\to_\P p_s\E\left[\left.\left(\mu_\lambda(s,X_i)-\mu_\lambda(0,X_i)-\E[\bar{\tau}_i^{\mathsf{post},s}(\lambda)|t_i^*=s]\right)^2\right\vert t_i^*=s\right]\\
&+p_s\E[\sigma_\lambda^2(s,X_i)|t_i^*=s]+\frac{1}{M}\E\left[e_s(X_i)\sigma_\lambda^2(0,X_i)\right]+\frac{\alpha(M,q)}{M^2}\E\left[\frac{e_s(X_i)^2}{e_\infty(X_i)}\sigma_\lambda^2(0,X_i)\right]
\end{align*}
by Lemma \ref{lemma_app:conv}. On the other hand, under finite moments, for $1\le i\le n$,
\begin{align*}
\E[Z_{n,i}^4|\mathcal{F}_{n,i}]&=\E\left[\left.\left(\frac{1}{\sqrt{n}}\1(t_i^*=s)\left(\mu_\lambda(s,X_i)-\mu_\lambda(0,X_i)-\E[\bar{\tau}_i^{\mathsf{post},s}(\lambda)|t_i^*=s]\right)\right)^4\right\vert\tstar,X_1,\ldots,X_{i-1}\right]\\
&=\frac{\1(t_i^*=s)}{n^2}\E\left[\left.\left(\mu_\lambda(s,X_i)-\mu_\lambda(s',X_i)-\E[\bar{\tau}_i^{\mathsf{post},s}(\lambda)|t_i^*=s]\right)^4\right\vert t_i^*=s\right]\\
&\le \frac{\1(t_i^*=s)}{n^2}C
\end{align*}
for some finite constant $C$ and for $n+1\le i\le 2n$,
\begin{align*}
\E[Z_{n,i}^4|\mathcal{F}_{n,i}]&=\E\left[\left.\left(\frac{1}{\sqrt{n}}\left(\1(t_{i-n}^*=s)-\1(t_{i-n}^*=\infty)\frac{K_M(i-n,s)}{M}\right)\varepsilon_{i-n}(t_{i-n}^*,X_{i-n})\right)^4\right\vert\mathcal{F}_{n,i-1}\right]\\
&=\frac{1}{n^2}\left(\1(t_{i-n}^*=s)-\1(t_{i-n}^*=s')\frac{K_M(i-n,s)}{M}\right)^4\E\left[\left.\varepsilon^4_{i-n}(t_{i-n}^*,X_{i-n})\right\vert t_{i-n}^*,X_{i-n}\right]\\
&=\frac{\1(t_{i-n}^*=s)}{n^2}\E\left[\left.\varepsilon^4_{i-n}(t_{i-n}^*,X_{i-n})\right\vert t_{i-n}^*=s,X_{i-n}\right]\\
&+\frac{\1(t_{i-n}^*=\infty)}{n^2}\cdot\frac{K_M(i-n,s)^4}{M^2}\E\left[\left.\varepsilon^4_{i-n}(t_{i-n}^*,X_{i-n})\right\vert t_{i-n}^*=\infty,X_{i-n}\right]\\
&\le C'\left(\frac{\1(t_{i-n}^*=s)}{n^2}+\frac{\1(t_{i-n}^*=\infty)}{n^2}\cdot\frac{K_M(i-n,s)^4}{M^2}\right)
\end{align*}
for some finite constant $C'$. Thus,
\begin{align*}
\sum_{i=1}^{2n}\E[Z_{n,i}^4|\mathcal{F}_{n,i}]&\le \frac{C}{n}\cdot \frac{1}{n}\sum_{i=1}^n \1(t_i^*=s)+\frac{C'}{n}\left(\frac{1}{n}\sum_{i=n+1}^{2n} \1(t_{i-n}^*=s)+\frac{1}{n}\sum_{i=n+1}^{2n}\frac{\1(t_{i-n}^*=\infty)K_M(i-n,s)^4}{M^2}\right)\\
&\to_\P 0
\end{align*}
because by Lemma 3 in \citet{Abadie-Imbens_2006_ECMA} $\E[K_M(i,s)^4]$ is uniformly bounded. Thus, by the martingale central limit theorem and the Slutsky theorem,
\begin{align*}
\sqrt{n}\left(\hat\tau(w^s,\lambda)-\E[\bar{\tau}_i^{\mathsf{post},s}(\lambda)|t_i^*=s]-B_\lambda(s,M)\right)\to_\mathcal{D}\mathcal{N}(0,V(s,\lambda))
\end{align*}
where
\begin{align*}
V(s,\lambda)&= \frac{1}{p_s}\E\left[\left.\left(\mu_\lambda(s,X_i)-\mu_\lambda(0,X_i)-\E[\bar{\tau}_i^{\mathsf{post},s}(\lambda)|t_i^*=s]\right)^2\right\vert t_i^*=s\right]+\frac{1}{p_s}\E[\sigma_\lambda^2(s,X_i)|t_i^*=s]\\
&+\frac{1}{p_s^2M^2}\E\left[\left(M+\alpha(M,q)\frac{e_s(X_i)}{e_\infty(X_i)}\right)e_s(X_i)\sigma_\lambda^2(0,X_i)\right]
\end{align*}
which completes the proof. $\square$


\subsection*{Proof of Proposition \ref{prop:seb}}

From Theorem \ref{thm:asynorm},
\begin{align*}
V(s,\lambda)&= \frac{1}{p_s}\E\left[\left.\left(\E[\bar{\tau}_i^{\mathsf{post},s}(\lambda)|t_i^*=s,X_i]-\E[\bar{\tau}_i^{\mathsf{post},s}(\lambda)|t_i^*=s]\right)^2\right\vert t_i^*=s\right]+\frac{1}{p_s}\E[\sigma_\lambda^2(s,X_i)|t_i^*=s]\\
&+\frac{1}{p_s^2M^2}\E\left[\left(M+\alpha(M,q)\frac{e_s(X_i)}{e_\infty(X_i)}\right)e_s(X_i)\sigma_\lambda^2(0,X_i)\right]
\end{align*}
but
\begin{multline*}
\frac{1}{p_s}\E\left[\left.\left(\E[\bar{\tau}_i^{\mathsf{post},s}(\lambda)|t_i^*=s,X_i]-\E[\bar{\tau}_i^{\mathsf{post},s}(\lambda)|t_i^*=s]\right)^2\right\vert t_i^*=s\right]=\\
\frac{1}{p_s^2}\E\left[\left(\E[\bar{\tau}_i^{\mathsf{post},s}(\lambda)|t_i^*=s,X_i]-\E[\bar{\tau}_i^{\mathsf{post},s}(\lambda)|t_i^*=s]\right)^2e_s(X_i)\right]
\end{multline*}
by the law of iterated expectations. Similarly,
\begin{align*}
\frac{1}{p_s}\E\left[\sigma^2_\lambda(s,X_i)|t_i^*=s\right]&=\frac{1}{p_s^2}\E\left[\sigma^2_\lambda(s,X_i)e_s(X_i)\right]
\end{align*}
and the result follows from straightforward algebra. $\square$


\subsection*{Proof of Proposition \ref{prop:cluster_se}}

The regression residual is:
\begin{align*}
\hat\varepsilon_i&=\Delta \bar{Y}_i^s(\lambda)-\left(\frac{1}{N_s}\sum_j\1(t_j^*=s)\Delta \bar{Y}_j^s(\lambda)\right)\1(t_i^*=s)\\
&-\left(\frac{1}{N_s}\sum_j\1(t_j^*=\infty)\frac{K_M(j,s)}{M}\Delta \bar{Y}_j^s(\lambda)\right)\1(t_i^*=\infty)\\
&=\Delta \bar{Y}_i^s(\lambda)-W_i'\hat\mu
\end{align*}
where
\[W_i=(\1(t_i^*=s),\1(t_i^*=\infty))',\quad \hat\mu=\left(\frac{1}{N_s}\sum_j\1(t_j^*=s)\Delta \bar{Y}_j^s(\lambda),\frac{1}{N_s}\sum_j\1(t_j^*=\infty)\frac{K_M(j,s)}{M}\Delta \bar{Y}_j^s(\lambda)\right)'.\]
Start by noting that by the law of large numbers and Lemma \ref{lemma_app:conv}, 
\[\hat\mu\to_\P \mu:=(\E[\Delta \bar{Y}_i^s(\lambda)|t_i^*=s],\E\left[\E[\Delta \bar{Y}_i^s(\lambda)|t_i^*=\infty,X_i]|t_i^*=s\right])'.\]
Define $\varepsilon_i=\Delta \bar{Y}_i^s(\lambda)-W_i'\mu$ and note that $\hat\varepsilon_i=\varepsilon_i-W_i'(\hat\mu-\mu)$. By the triangle and Schwarz inequalities,
\begin{align*}
\abs{\hat\varepsilon_i^2-\varepsilon_i^2}\le \norm{W_i}\norm{\hat\mu-\mu}^2+2\abs{\varepsilon_i}\norm{W_i}\norm{\hat\mu-\mu}\le \norm{\hat\mu-\mu}^2+2\abs{\varepsilon_i}\norm{\hat\mu-\mu}
\end{align*}
using that $\norm{W_i}=\1(t_i^*=s)+\1(t_i^*=\infty)\le 1$. Then we have that:
\begin{align*}
n\hat{V}_\mathsf{cr}&=\frac{n}{N_s^2}\sum_i\1(t_i^*=s)\hat\varepsilon_i^2+\frac{n}{N_s^2}\sum_i\1(t_i^*=\infty)\frac{K_M(i,s)^2}{M^2}\hat\varepsilon_i^2\\
&=\frac{n^2}{N_s^2}\left\{\frac{1}{n}\sum_i\1(t_i^*=s)\varepsilon_i^2+\frac{1}{n}\sum_i\1(t_i^*=\infty)\frac{K_M(i,s)^2}{M^2}\varepsilon_i^2\right\}\\
&+\frac{n^2}{N_s^2}\cdot\frac{1}{n}\sum_i\left(\1(t_i^*=s)-\1(t_i^*=\infty)\frac{K_M(i,s)}{M}\right)^2(\hat\varepsilon_i^2-\varepsilon_i^2).
\end{align*}
Define
\[R_n=\frac{1}{n}\sum_i\left(\1(t_i^*=s)-\1(t_i^*=\infty)\frac{K_M(i,s)}{M}\right)^2(\hat\varepsilon_i^2-\varepsilon_i^2).\]
We have that:
\begin{align*}
\abs{R_n}&\le \frac{1}{n}\sum_i\left(\1(t_i^*=s)-\1(t_i^*=\infty)\frac{K_M(i,s)}{M}\right)^2\norm{\hat\mu-\mu}^2\\
&+\frac{2}{n}\sum_i\left(\1(t_i^*=s)-\1(t_i^*=\infty)\frac{K_M(i,s)}{M}\right)^2\abs{\varepsilon_i}\norm{\hat\mu-\mu}\\
:&=R_{1,n}\norm{\hat\mu-\mu}^2+2R_{2,n}\norm{\hat\mu-\mu}.
\end{align*}
Note that
\[\E[R_{1,n}]\le 1+\frac{1}{M^2}\E\left[K_M(i,s)^2\right]<C\]
by Lemma 3 in \citet{Abadie-Imbens_2006_ECMA} and thus by Markov's inequality $R_{1,n}=O_\P(1)$ and $R_{1,n}\norm{\hat\mu-\mu}^2=o_\P(1)$. Similarly, for the second term,
\begin{align*}
\E[R_{2,n}]&=\E\left[\left(\1(t_i^*=s)-\1(t_i^*=\infty)\frac{K_M(i,s)}{M}\right)^2\abs{\varepsilon_i}\right]\\
&\le \left(1+\frac{\E\left[K_M(i,s)^4\right]}{M^4}\right)^{1/2}\E[\varepsilon_i^2]^{1/2}
\end{align*}
which again is uniformly bounded and thus $2R_{2,n}\norm{\hat\mu-\mu}=o_\P(1)$. Then,
\begin{align*}
n\hat{V}_\mathsf{cr}&=\frac{n^2}{N_s^2}\left\{\frac{1}{n}\sum_i\1(t_i^*=s)\varepsilon_i^2+\frac{1}{n}\sum_i\1(t_i^*=\infty)\frac{K_M(i,s)^2}{M^2}\varepsilon_i^2\right\}+o_\P(1)\\
:&=\frac{n^2}{N_s^2}\left\{T_{1,n}+T_{2,n}\right\}+o_\P(1)
\end{align*}
By the law of large numbers, $T_{1,n}\to_\P \E[\varepsilon_i^2|t_i^*=s]p_s$. On the other hand, by Lemma \ref{lemma_app:conv},
\begin{align*}
T_{2,n}\to_\P \frac{1}{M}\E\left[e_s(X_i)\E[\varepsilon_i^2|t_i^*=\infty,X_i]\right]+\frac{\alpha(M,q)}{M^2}\E\left[\frac{e_s(X_i)^2}{e_\infty(X_i)}\E[\varepsilon_i^2|t_i^*=\infty,X_i]\right].
\end{align*}
Collecting the results,
\begin{align*}
n\hat{V}_\mathsf{cr}&\to_\P \frac{\E[\varepsilon_i^2|t_i^*=s]}{p_s}+\frac{1}{Mp_s^2}\E\left[e_s(X_i)\E[\varepsilon_i^2|t_i^*=\infty,X_i]\right]+\frac{\alpha(M,q)}{M^2p_s^2}\E\left[\frac{e_s(X_i)^2}{e_\infty(X_i)}\E[\varepsilon_i^2|t_i^*=\infty,X_i]\right]\\
&=\frac{\E[\varepsilon_i^2|t_i^*=s]}{p_s}+\frac{1}{p_s^2M^2}\E\left[\left(M+\alpha(M,q)\frac{e_s(X_i)}{e_\infty(X_i)}\right)e_s(X_i)\E[\varepsilon_i^2|t_i^*=\infty,X_i]\right].
\end{align*}
Now,
\begin{align*}
\E[\varepsilon_i^2|t_i^*=s]&=\E[(\Delta \bar{Y}_i^s(\lambda)-\E[\Delta \bar{Y}_i^s(\lambda)|t_i^*=s])^2|t_i^*=s]=\V[\Delta \bar{Y}_i^s(\lambda)|t_i^*=s]\\
&=\V[\mu_\lambda(s,X_i)|t_i^*=s]+\E[\sigma^2_\lambda(s,X_i)|t_i^*=s]\\
&=\V[\E[\bar{\tau}_i^{\mathsf{post},s}(\lambda)|t_i^*=s,X_i]|t_i^*=s]+\V[\E[\Delta \bar{Y}_i^s(0,\lambda)|t_i^*=s,X_i]|t_i^*=s]\\
&+2\cov(\E[\bar{\tau}_i^{\mathsf{post},s}(\lambda)|t_i^*=s,X_i],\E[\Delta \bar{Y}_i^s(0,\lambda)|t_i^*=s,X_i]|t_i^*=s)+\E[\sigma^2_\lambda(s,X_i)|t_i^*=s]
\end{align*}
and
\begin{align*}
\E[\varepsilon_i^2|t_i^*=\infty,X_i]&=\E[(\Delta \bar{Y}_i^s(\lambda)-\E[\mu(0,X_i)|t_i^*=s])^2|t_i^*=\infty,X_i]\\
&=\E[(\Delta \bar{Y}_i^s(\lambda)-\mu_\lambda(0,X_i))^2|t_i^*=\infty,X_i]+\E[(\mu_\lambda(0,X_i)-\E[\mu_\lambda(0,X_i)|t_i^*=s])^2|t_i^*=\infty,X_i]\\
&=\sigma^2_\lambda(0,X_i)+(\mu_\lambda(0,X_i)-\E[\mu(0,X_i)|t_i^*=s])^2
\end{align*}
Next, use that under conditional parallel trends,
\begin{align*}
\mu_\lambda(0,X_i)&=\E[\Delta \bar{Y}_i^s(0,\lambda)|t_i^*=s,X_i]\\
\E[\mu_\lambda(0,X_i)|t_i^*=s]&=\E[\Delta \bar{Y}_i^s(0,\lambda)|t_i^*=s]
\end{align*}
Collecting the results,
\begin{align*}
n\hat{V}_\mathsf{cr}&\to_\P V(s,\lambda)\\
&+\frac{1}{p_s}\V[\E[\Delta\bar{Y}_i^s(0,\lambda)|t_i^*=s,X_i]|t_i^*=s]\\
&+\frac{2}{p_s}\cov(\E[\bar{\tau}_i^{\mathsf{post},s}(\lambda)|t_i^*=s,X_i],\E[\Delta\bar{Y}_i^s(0,\lambda)|t_i^*=s,X_i]|t_i^*=s)\\
&+\frac{1}{p_s^2M^2}\E\left[\left(M+\alpha(M,q)\frac{e_s(X_i)}{e_\infty(X_i)}\right)e_s(X_i)(\E[\Delta\bar{Y}_i^s(0,\lambda)|t_i^*=s,X_i]-\E[\Delta\bar{Y}_i^s(0,\lambda)|t_i^*=s])^2\right].
\end{align*}
as required. $\square$


\subsection*{Proof of Proposition \ref{prop:discrete}}

Suppose the vector $X_i$ includes only discrete variables. By Lemma \ref{lemma:identif},
\begin{align*}
\E[\bar\tau_i^{\mathsf{post},s}(\lambda)|t_i^*=s]&=\E[\Delta \bar{Y}_i^s(\lambda)|t_i^*=s]-\sum_{x\in\mathcal{X}}\mu_\lambda(0,x)\P[X_i=x|t_i^*=s]\\
&=\E[\Delta \bar{Y}_i^s(\lambda)|t_i^*=s]-\sum_{x\in\mathcal{X}}\mu_\lambda(0,x)\P[t_i^*=s|X_i=x]\frac{p(x)}{p_s}\\
&=\E[\Delta \bar{Y}_i^s(\lambda)|t_i^*=s]-\sum_{x\in\mathcal{X}}\mu_\lambda(0,x)e_s(x)\frac{p(x)}{p_s}
\end{align*}
where $e_s(x)=\P[t_i^*=s|X_i=x]$, $p(x)=\P[X_i=x]$ and $p_s=\P[t_i^*=s]$. This term can be estimated as:
\begin{align*}
\hat\tau(s,\lambda)&=\frac{1}{N_s}\sum_i\Delta \bar{Y}_i^s(\lambda)\1(t_i^*=s)-\sum_{x\in\mathcal{X}}\frac{\sum_i\Delta \bar{Y}_i^s(\lambda)\1(t_i^*=\infty)\1(X_i=x)}{\sum_i\1(t_i^*=\infty)\1(X_i=x)}\cdot \frac{\sum_i\1(t_i^*=s)\1(X_i=x)}{\sum_i\1(t_i^*=s)}\\
&=\frac{1}{N_s}\sum_i\Delta \bar{Y}_i^s(\lambda)\1(t_i^*=s)-\sum_{x\in\mathcal{X}}\frac{\sum_i\Delta \bar{Y}_i^s(\lambda)\1(t_i^*=\infty)\1(X_i=x)}{N_s}\cdot \frac{\hat{e}_s(x)}{\hat{e}_\infty(x)}\\
&=\frac{1}{N_s}\sum_i\Delta \bar{Y}_i^s(\lambda)\1(t_i^*=s)-\frac{1}{N_s}\sum_i\Delta \bar{Y}_i^s(\lambda)\1(t_i^*=\infty)\sum_{x\in\mathcal{X}}\1(X_i=x) \frac{\hat{e}_s(x)}{\hat{e}_\infty(x)}\\
&=\frac{1}{N_s}\sum_i\left(\1(t_i^*=s)-\1(t_i^*=\infty)\frac{\hat{e}_s(X_i)}{\hat{e}_\infty(X_i)}\right)\Delta \bar{Y}_i^s(\lambda).
\end{align*}
Following the proof of Theorem \ref{thm:asynorm}, the estimator can be rewritten as:
\begin{align*}
\hat\tau(s,\lambda)&=\frac{1}{N_s}\sum_i\1(t_i^*=s)(\mu(s,X_i)-\mu(0,X_i))\\
&+\frac{1}{N_s}\sum_i\left(\1(t_i^*=s)-\1(t_i^*=\infty)\frac{\hat{e}_s(X_i)}{\hat{e}_\infty(X_i)}\right)(\Delta \bar{Y}_i^s(\lambda)-\mu_\lambda(t_i^*,X_i))\\
&+B_\lambda(s)
\end{align*}
where in this case:
\begin{align*}
B_\lambda(s)&=\frac{1}{N_s}\sum_i\left(\1(t_i^*=s)-\1(t_i^*=\infty)\frac{\hat{e}_s(X_i)}{\hat{e}_\infty(X_i)}\right)\mu_\lambda(0,X_i).
\end{align*}
When covariates are discrete, the matching can be done exactly, and thus the bias is zero. To see this,
\begin{align*}
B_\lambda(s)&=\frac{1}{N_s}\sum_i\left(\1(t_i^*=s)-\1(t_i^*=\infty)\frac{\hat{e}_s(X_i)}{\hat{e}_\infty(X_i)}\right)\mu_\lambda(0,X_i)\\
&=\sum_{x\in\mathcal{X}}\mu_\lambda(0,x)\frac{1}{N_s}\sum_i\left(\1(t_i^*=s)-\1(t_i^*=\infty)\frac{\hat{e}_s(x)}{\hat{e}_\infty(x)}\right)\1(X_i=x)\\
&=\sum_{x\in\mathcal{X}}\mu_\lambda(0,x)\left(\frac{\sum_i\1(t_i^*=s)\1(X_i=x)}{\sum_i\1(t_i^*=s)}-\frac{\sum_i\1(t_i^*=s)\1(X_i=x)}{\sum_i\1(t_i^*=\infty)\1(X_i=x)}\cdot\frac{\sum_i\1(t_i^*=\infty)\1(X_i=x)}{\sum_i\1(t_i^*=s)}\right)\\
&=0.
\end{align*}
Thus,
\begin{align*}
\hat\tau(s,\lambda)-\E[\bar{\tau}_i^{\mathsf{post},s}(\lambda)|t_i^*=s]&=\frac{1}{N_s}\sum_i\1(t_i^*=s)(\mu_\lambda(s,X_i)-\mu_\lambda(0,X_i)-\E[\bar{\tau}_i^{\mathsf{post},s}(\lambda)|t_i^*=s])\\
&+\frac{1}{N_s}\sum_i\left(\1(t_i^*=s)-\1(t_i^*=\infty)\frac{e_s(X_i)}{e_\infty(X_i)}\right)(\Delta \bar{Y}_i^s(\lambda)-\mu_\lambda(t_i^*,X_i))\\
&-\sum_{x\in\mathcal{X}}\left(\frac{\hat{e}_s(x)}{\hat{e}_\infty(x)}-\frac{e_s(x)}{e_\infty(x)}\right)\frac{1}{N_s}\sum_i\1(t_i^*=\infty)\1(X_i=x)(\Delta \bar{Y}_i^s(\lambda)-\mu_\lambda(0,x)).
\end{align*}
Now,
\begin{align*}
\frac{1}{N_s}\sum_i\1(t_i^*=\infty)\1(X_i=x)(\Delta \bar{Y}_i^s(\lambda)-\mu_\lambda(0,x))&\to_\P \frac{e_\infty(x)p_x}{p_s}\E\left[\left.\Delta \bar{Y}_i^s(\lambda)-\mu_\lambda(0,x)\right\vert t_i^*=\infty,X_i=x\right]\\
&=0
\end{align*}
under finite conditional moments of the outcome whereas
\begin{align*}
\frac{\hat{e}_s(x)}{\hat{e}_\infty(x)}-\frac{e_s(x)}{e_\infty(x)}&=\frac{\hat{e}_s(x)e_\infty(x)-\hat{e}_\infty(x)e_s(x)}{\hat{e}_\infty(x)e_\infty(x)}\\
&=\frac{(\hat{e}_s(x)-e_s(x))e_\infty(x)-(\hat{e}_\infty(x)-e_\infty(x))e_s(x)}{\hat{e}_\infty(x)e_\infty(x)}\\
&=\frac{1}{\hat{e}_\infty(x)}(\hat{e}_s(x)-e_s(x))-\frac{1}{\hat{e}_\infty(x) }\frac{e_s(x)}{e_\infty(x)}(\hat{e}_\infty(x)-e_\infty(x))
\end{align*}
and
\begin{align*}
\hat{e}_s(x)-e_s(x)&=\frac{\sum_i (\1(t_i^*=s)-e_s(x))\1(X_i=x)}{\sum_i\1(X_i=x)}=\frac{1}{n}\sum_i\frac{ (\1(t_i^*=s)-e_s(x))\1(X_i=x)}{\hat{p}(x)}
\end{align*}
from which
\begin{align*}
\frac{\hat{e}_s(x)}{\hat{e}_\infty(x)}-\frac{e_s(x)}{e_\infty(x)}&=\frac{1}{\hat{e}_\infty(x)\hat{p}(x)}\cdot\frac{1}{n}\sum_i(\1(t_i^*=s)-e_s(x))\1(X_i=x)\\
&-\frac{1}{\hat{e}_\infty(x)\hat{p}(x)}\frac{e_s(x)}{e_\infty(x)}\cdot\frac{1}{n}\sum_i(\1(t_i^*=\infty)-e_\infty(x))\1(X_i=x)\\
&=O_\P(n^{-1/2})
\end{align*}
as long as $\hat{e}_s(x)>0$ for all $s,x$. Therefore,
\[\sum_{x\in\mathcal{X}}\left(\frac{\hat{e}_s(x)}{\hat{e}_\infty(x)}-\frac{e_s(x)}{e_\infty(x)}\right)\frac{1}{N_s}\sum_i\1(t_i^*=\infty)\1(X_i=x)(\Delta \bar{Y}_i^s(\lambda)-\mu_\lambda(0,x))=o_\P(n^{-1/2})\]
and
\begin{align*}
\sqrt{n}(\hat\tau(s,\lambda)-\E[\bar{\tau}_i^{\mathsf{post},s}(\lambda)|t_i^*=s])&=\frac{1}{\sqrt{n}}\sum_i\psi_i(s,\lambda)+o_\P(1)\\
&\to_\mathcal{D}\mathcal{N}(0,\V[\psi_i(s,\lambda)])
\end{align*}
where
\begin{align*}
\psi_i(s,\lambda)&=\frac{\1(t_i^*=s)}{p_s}(\mu_\lambda(s,X_i)-\mu_\lambda(0,X_i)-\E[\bar{\tau}_i^{\mathsf{post},s}(\lambda)|t_i^*=s])\\
&+\frac{1}{p_s}\left(\1(t_i^*=s)-\1(t_i^*=\infty)\frac{e_s(X_i)}{e_\infty(X_i)}\right)(\Delta \bar{Y}_i^s(\lambda)-\mu_\lambda(t_i^*,X_i)).
\end{align*}
The variance can be expressed as:
\begin{align*}
\V[\psi_i(s,\lambda)]&=\frac{1}{p_s}\E\left[\left.(\E[\bar{\tau}_i^{\mathsf{post},s}(\lambda)|t_i^*=s,X_i]-\E[\bar{\tau}_i^{\mathsf{post},s}(\lambda)|t_i^*=s])^2\right\vert t_i^*=s\right]+\frac{1}{p_s}\E[\sigma^2_\lambda(s,X_i)|t_i^*=s]\\
&+\frac{1}{p_s^2}\E\left[\frac{e_s(X_i)^2}{e_\infty(X_i)}\sigma^2(0,X_i)\right]
\end{align*}
as required. $\square$


\subsection*{Proof of Proposition \ref{prop:wo_repl}}

By Proposition \ref{prop:weighted_reg}, the estimator of interest can be written as the estimator from a regression of $\Delta Y_i^s(\lambda)$ on an intercept and a treated cohort indicator $\1(t_i^*=s)$. This corresponds to the case analyzed in \citet{Abadie-Spiess_2022_JASA} with $Z_i=(1,\1(t_i^*=s))'$. Under the conditions in the proposition's statement, Assumptions 1-5 in  \citet{Abadie-Spiess_2022_JASA} hold, and thus their Proposition 2 apply to $\hat\beta=(\hat\alpha,\hat\tau_\mathsf{nr}(\lambda))'$ where $\hat\alpha$ is the intercept from the regression. Thus,
\[\sqrt{(M+1)N_s}(\hat\beta-\beta)\to_\mathcal{D}\mathcal{N}(0,H^{-1}JH^{-1})\]
where in this case $\beta=(\E[\mu_\lambda(0,X_i)|t_i^*=s],\E[\bar{\tau}_i^{\mathsf{post},s}(\lambda)|t_i^*=s])'$,
\begin{align*}
H=\frac{1}{1+M}\begin{bmatrix}
1+M & 1\\
1 & 1
\end{bmatrix},\quad H^{-1}=\frac{1+M}{M}\begin{bmatrix}
1 & -1\\
-1 & 1+M
\end{bmatrix}
\end{align*}
and $J=J_1+J_2$ where
\begin{align*}
J_1&=\frac{1}{1+M}\V\left[\left.\begin{matrix}
\mu_\lambda(s,X_i)+M\mu_\lambda(0,X_i)\\
\mu_\lambda(s,X_i)
\end{matrix}\right\vert t_i^*=s\right],\\
J_2&=\frac{1}{1+M}\E\left[\left.\begin{matrix}
\sigma_\lambda^2(s,X_i)+M\sigma_\lambda^2(0,X_i) & \sigma_\lambda^2(s,X_i)\\
\sigma_\lambda^2(s,X_i) & \sigma_\lambda^2(s,X_i)
\end{matrix}\right\vert t_i^*=s\right].
\end{align*}
Then, by straightforward algebra,
\begin{align*}
V_\mathsf{nr}(\lambda)=H^{-1}JH^{-1}_{(2,2)}&=(1+M)\E\left[\left.\left(\E[\bar{\tau}_i^{\mathsf{post},s}(\lambda)|t_i^*=s,X_i]-\E[\bar{\tau}_i^{\mathsf{post},s}(\lambda)|t_i^*=s]\right)^2\right\vert t_i^*=s\right]\\
&+(1+M)\E\left[\left.\sigma^2_\lambda(s,X_i)+\frac{\sigma^2_\lambda(0,X_i)}{M}\right\vert t_i^*=s\right]
\end{align*}
as required. $\square$


\section{Proofs of Auxiliary Results}

\subsection*{Proof of Lemma \ref{lemma_app:2wfe}}

Let $\ddot D_{it}=D_{it}-\bar{D}_i-\tilde{D}_t+\bar{D}$ and note that since $D_{it}=\1(t\ge t_i^*)=\sum_{s<\infty}\1(t_i^*=s)\1(t\ge s)$,
\begin{align*}
\bar{D}_i&=\sum_{s<\infty}\1(t_i^*=s)\frac{\sum_t \lambda_t\1(t\ge s)}{\sum_t\lambda_t}=\sum_{s<\infty}\1(t_i^*=s)\Lambda_s\\
\tilde{D}_t&=\sum_{s<\infty}\1(t\ge s)\frac{\sum_iw_i\1(t_i^*=s)}{\sum_iw_i}=\sum_{s<\infty}\1(t\ge s)\hat{p}_s^w\\
\bar{D}&=\sum_{s<\infty}\hat{p}_s^w\Lambda_s
\end{align*}
where $\Lambda_s=\sum_t\lambda_t\1(t\ge s)/\sum_t\lambda_t$ is the proportion of included post-periods $t\ge s$ and $\hat{p}_s^w=\sum_iw_i\1(t_i^*=s)/\sum_iw_i$ is the weighted proportion of cohort $s$ in the sample. This implies that
\begin{align*}
\ddot{D}_{it}&=\sum_{s<\infty}(\1(t_i^*=s)-\hat{p}_s^w)(\1(t\ge s)-\Lambda_s)
\end{align*}
and
\begin{align*}
D_{it}\ddot{D}_{it}&=\sum_{s<\infty}(\1(t_i^*=s)-\hat{p}_s^w)(\1(t\ge s)-\Lambda_s)\sum_{s'<\infty}\1(t_i^*=s')\1(t\ge s')\\
&=\sum_{s<\infty}(\1(t_i^*=s)-\hat{p}_s^w)(\1(t\ge s)-\Lambda_s)\1(t_i^*=s)\1(t\ge s)\\
&+\sum_{s<\infty}\sum_{s'\ne s,\infty}(\1(t_i^*=s)-\hat{p}_s^w)(\1(t\ge s)-\Lambda_s)\1(t_i^*=s')\1(t\ge s')\\
&=\sum_{s<\infty}(1-\hat{p}_s^w)(1-\Lambda_s)\1(t_i^*=s)\1(t\ge s)\\
&-\sum_{s<\infty}\sum_{s'\ne s,\infty}\hat{p}_s^w(\1(t\ge s)-\Lambda_s)\1(t_i^*=s')\1(t\ge s')
\end{align*}
from which:
\begin{align*}
\sum_i\sum_tw_i\lambda_tD_{it}\ddot{D}_{it}&=\sum_i\sum_t\sum_{s<\infty}w_i\lambda_t(1-\hat{p}_s^w)(1-\Lambda_s)\1(t_i^*=s)\1(t\ge s)\\
&-\sum_i\sum_t\sum_{s<\infty}\sum_{s'\ne s,\infty}w_i\lambda_t\hat{p}_s^w(\1(t\ge s)-\Lambda_s)\1(t_i^*=s')\1(t\ge s')\\
&=\left(\sum_i\sum_tw_i\lambda_t\right)\left(\sum_{s<\infty}\hat{p}_s^w(1-\hat{p}_s^w)\Lambda_s(1-\Lambda_s)-\sum_{s<\infty}\sum_{s'\ne s,\infty}\hat{p}_s^w\hat{p}_{s'}^w(\Lambda_{s\vee s'}-\Lambda_s\Lambda_{s'})\right)
\end{align*}
where $s\vee s'=\max\{s,s'\}$. Using that $1-\hat{p}_s^w=\sum_{s'\ne s}\hat{p}_{s'}^w=\hat{p}_\infty^w+\sum_{s'\ne s,\infty}\hat{p}_{s'}^w$, this term can be rewritten as:
\begin{align*}
\sum_i\sum_tw_i\lambda_tD_{it}\ddot{D}_{it}&=\sum_{s<\infty}\hat{p}_s^w\left\{\hat{p}_\infty^w\Lambda_s(1-\Lambda_s)+\sum_{s'<s}\hat{p}_{s'}^w\Lambda_s(\Lambda_{s'}-\Lambda_s)+\sum_{s<s'<\infty}\hat{p}_{s'}^w(1-\Lambda_s)(\Lambda_s-\Lambda_{s'})\right\}\\
&\times \left(\sum_iw_i\sum_t\lambda_t\right).
\end{align*}
Next, let
\[\bar{Y}_i^{\mathsf{post},s}(\lambda)=\frac{\sum_t\lambda_t\1(t\ge s)Y_{it}}{\sum_t\lambda_t\1(t\ge s)},\quad \bar{Y}_i^{\mathsf{pre},s}(\lambda)=\frac{\sum_t\lambda_t\1(t<s)Y_{it}}{\sum_t\lambda_t\1(t<s)}.\]
We have that:

\begin{align*}
\sum_i\sum_tw_i\lambda_tY_{it}\ddot D_{it}&=\sum_i\sum_t\sum_{s<\infty}w_i\lambda_tY_{it}(\1(t_i^*=s)-\hat{p}_s^w)(\1(t\ge s)-\Lambda_s)\\
&=\sum_i\sum_{s<\infty}w_i(\1(t_i^*=s)-\hat{p}_s^w)\sum_t\lambda_tY_{it}(\1(t\ge s)-\Lambda_s)\\
&=\sum_{s<\infty}\sum_iw_i(\1(t_i^*=s)-\hat{p}_s^w)\Lambda_s(1-\Lambda_s)(\bar{Y}_i^{\mathsf{post},s}(\lambda)-\bar{Y}_i^{\mathsf{pre},s}(\lambda))\left(\sum_t\lambda_t\right)\\
&=\sum_{s<\infty}(1-\hat{p}_s^w)\Lambda_s(1-\Lambda_s)\sum_iw_i\1(t_i^*=s)(\bar{Y}_i^{\mathsf{post},s}(\lambda)-\bar{Y}_i^{\mathsf{pre},s}(\lambda))\left(\sum_t\lambda_t\right)\\
&-\sum_{s<\infty}\hat{p}_s^w\Lambda_s(1-\Lambda_s)\sum_iw_i\1(t_i^*\ne s)(\bar{Y}_i^{\mathsf{post},s}(\lambda)-\bar{Y}_i^{\mathsf{pre},s}(\lambda))\left(\sum_t\lambda_t\right)\\
&=\sum_{s<\infty}\hat{p}_s^w(1-\hat{p}_s^w)\Lambda_s(1-\Lambda_s)\frac{\sum_iw_i\1(t_i^*=s)(\bar{Y}_i^{\mathsf{post},s}(\lambda)-\bar{Y}_i^{\mathsf{pre},s}(\lambda))}{\sum_iw_i\1(t_i^*=s)}\left(\sum_i\sum_tw_i\lambda_t\right)\\
&-\sum_{s<\infty}\hat{p}_s^w(1-\hat{p}_s^w)\Lambda_s(1-\Lambda_s)\frac{\sum_iw_i\1(t_i^*\ne s)(\bar{Y}_i^{\mathsf{post},s}(\lambda)-\bar{Y}_i^{\mathsf{pre},s}(\lambda))}{\sum_iw_i\1(t_i^*\ne s)}\left(\sum_i\sum_tw_i\lambda_t\right)\\
&=\left(\sum_i\sum_tw_i\lambda_t\right)\sum_{s<\infty}\hat{p}_s^w(1-\hat{p}_s^w)\Lambda_s(1-\Lambda_s)\\
&\quad \times \left\{\frac{\sum_iw_i\1(t_i^*=s)(\bar{Y}_i^{\mathsf{post},s}(\lambda)-\bar{Y}_i^{\mathsf{pre},s}(\lambda))}{\sum_iw_i\1(t_i^*=s)}-\frac{\sum_iw_i\1(t_i^*\ne s)(\bar{Y}_i^{\mathsf{post},s}(\lambda)-\bar{Y}_i^{\mathsf{pre},s}(\lambda))}{\sum_iw_i\1(t_i^*\ne s)}\right\}\\
&=\left(\sum_i\sum_tw_i\lambda_t\right)\sum_{s<\infty}\hat{p}_s^w(1-\hat{p}_s^w)\Lambda_s(1-\Lambda_s)\\
&\quad \times \left\{\frac{\sum_iw_i\1(t_i^*=s)(\bar{Y}_i^{\mathsf{post},s}(\lambda)-\bar{Y}_i^{\mathsf{pre},s}(\lambda))}{\sum_iw_i\1(t_i^*=s)}\right.\\
&\left.-\sum_{s'\ne s}\frac{\sum_iw_i\1(t_i^*= s')(\bar{Y}_i^{\mathsf{post},s}(\lambda)-\bar{Y}_i^{\mathsf{pre},s}(\lambda))}{\sum_iw_i\1(t_i^*=s')}\frac{\hat{p}_{s'}^w}{\sum_{s'\ne s}\hat{p}_{s'}^w}\right\}.
\end{align*}
Defining:
\[\overline{\Delta Y}_i^s(\lambda)=\bar{Y}_i^{\mathsf{post},s}(\lambda)-\bar{Y}_i^{\mathsf{pre},s}(\lambda)\]
and using that $1-\hat{p}_s^w=\sum_{s'\ne s}\hat{p}_{s'}^w$ we can rewrite this as:
\begin{align*}
\sum_i\sum_tw_i\lambda_tY_{it}\ddot D_{it}&=\sum_{s<\infty}\sum_{s'\ne s}\hat{p}_s^w\hat{p}_{s'}^w\Lambda_s(1-\Lambda_s)\left\{\frac{\sum_iw_i\1(t_i^*=s)\overline{\Delta Y}_i^s(\lambda)}{\sum_iw_i\1(t_i^*=s)}-\frac{\sum_iw_i\1(t_i^*= s')\overline{\Delta Y}_i^s(\lambda)}{\sum_iw_i\1(t_i^*=s')}\right\}\\
&\times \left(\sum_i\sum_tw_i\lambda_t\right).
\end{align*}
Finally note that:
\begin{align*}
\overline{\Delta Y}_i^s(\lambda)&=\frac{\sum_t\lambda_tY_{it}\1(t\ge s)}{\sum_t\lambda_t\1(t\ge s)}-\bar{Y}_i^{\mathsf{pre},s}(\lambda)=\frac{\sum_t \lambda_t(Y_{it}-\bar{Y}_i^{\mathsf{pre},s}(\lambda))\1(t\ge s)}{\sum_t\lambda_t\1(t\ge s)}\\
&=\frac{\sum_t\lambda_t\left(Y_{it}-\frac{\sum_{t'}\lambda_{t'}Y_{it'}\1(t'<s)}{\sum_{t'}\lambda_{t'}\1(t<s)}\right)\1(t\ge s)}{\sum_t\lambda_t\1(t\ge s)}\\
&=\frac{\sum_t\sum_{t'}\lambda_t\lambda_{t'}(Y_{it}-Y_{it'})\1(t\ge s)\1(t'<s)}{\sum_t\lambda_t\1(t\ge s)\sum_{t'}\lambda_{t'}\1(t'<s)}
\end{align*}
and thus
\begin{align*}
\frac{\sum_i w_i\1(t_i^*=s)\overline{\Delta Y}_i^s(\lambda)}{\sum_i w_i\1(t_i^*=s)}&=\frac{\sum_t\sum_{t'}\lambda_t\lambda_{t'}\frac{\sum_iw_i\1(t_i^*=s)(Y_{it}-Y_{it'})}{\sum_iw_i\1(t_i^*=s)}\1(t\ge s)\1(t'<s)}{\sum_t\lambda_t\1(t\ge s)\sum_{t'}\lambda_{t'}\1(t'<s)}
\end{align*}
from which 
\begin{align*}
\frac{\sum_i w_i\1(t_i^*=s)\overline{\Delta Y}_i^s(\lambda)}{\sum_i w_i\1(t_i^*=s)}-\frac{\sum_i w_i\1(t_i^*=s')\overline{\Delta Y}_i^s(\lambda)}{\sum_i w_i\1(t_i^*= s')}&=\frac{\sum_t\sum_{t'}\lambda_t\lambda_{t'}\hat\tau_{s'}^s(t,t')\1(t\ge s)\1(t'<s)}{\sum_t\lambda_t\1(t\ge s)\sum_{t'}\lambda_{t'}\1(t'<s)}
\end{align*}
where
\[\hat\tau_{s'}^s(t,t')=\frac{\sum_iw_i\1(t_i^*=s)(Y_{it}-Y_{it'})}{\sum_iw_i\1(t_i^*=s)}-\frac{\sum_iw_i\1(t_i^*=s')(Y_{it}-Y_{it'})}{\sum_iw_i\1(t_i^*=s')}.\]
Therefore,
\begin{align*}
\sum_i\sum_tw_i\lambda_tY_{it}\ddot D_{it}&=\sum_{s<\infty}\sum_{s'\ne s}\sum_{t\ge s}\sum_{t'<s}\frac{\lambda_t\lambda_{t'}\hat{p}_s^w\hat{p}_{s'}^w\Lambda_s(1-\Lambda_s)}{\sum_t\lambda_t\1(t\ge s)\sum_{t'}\lambda_{t'}\1(t'<s)}\hat\tau_{s'}^s(t,t')\\
&\times \left(\sum_i\sum_tw_i\lambda_t\right)\\
&=\left(\frac{\sum_iw_i}{\sum_t\lambda_t}\right)\sum_{s<\infty}\sum_{s'\ne s}\sum_{t\ge s}\sum_{t'<s}\lambda_t\lambda_{t'}\hat{p}_s^w\hat{p}_{s'}^w\hat\tau_{s'}^s(t,t').
\end{align*}
Collecting the results, we have that:
\begin{align*}
\hat\tau(w,\lambda)&=\frac{\hat\tau_\mathsf{num}(w,\lambda)}{\hat\tau_\mathsf{den}(w,\lambda)}
\end{align*}
where
\begin{align*}
\hat\tau_\mathsf{num}(w,\lambda)&=\sum_{s<\infty}\sum_{s'\ne s}\sum_{t\ge s}\sum_{t'<s}\lambda_t\lambda_{t'}\hat{p}_s^w\hat{p}_{s'}^w\hat\tau_{s'}^s(t,t')\\
\hat\tau_\mathsf{den}(w,\lambda)&=\left(\sum_t\lambda_t\right)^2\sum_{s<\infty}\hat{p}_s^w\left\{\hat{p}_\infty^w\Lambda_s(1-\Lambda_s)+\sum_{s'<s}\hat{p}_{s'}^w\Lambda_s(\Lambda_{s'}-\Lambda_s)+\sum_{s<s'<\infty}\hat{p}_{s'}^w(1-\Lambda_s)(\Lambda_s-\Lambda_{s'})\right\}.
\end{align*}
which completes the proof. $\square$

\subsection*{Proof of Lemma \ref{lemma_app:conv}}

Let $\tstar=(t_1^*,\ldots,t_n^*)$ and $\X=(X_1',\ldots,X_n')'$.
\begin{align*}
\frac{1}{N_s}\sum_i\1(t_i^*=s')\frac{K_M(i,s)}{M}Z_i&=\frac{1}{N_s}\sum_i\1(t_i^*=s)Z_i-\frac{1}{N_s}\sum_i\left(\1(t_i^*=s)-\1(t_i^*=s')\frac{K_M(i,s)}{M}\right)Z_i\\
&=\E[Z_i|t_i^*=s]+o_\P(1)-T_n
\end{align*}
where
\[T_n=\frac{1}{N_s}\sum_i\left(\1(t_i^*=s)-\1(t_i^*=s')\frac{K_M(i,s)}{M}\right)Z_i.\]
Now,
\begin{align*}
T_n=\frac{n}{N_s}&\left\{\frac{1}{n}\sum_i\left(\1(t_i^*=s)-\1(t_i^*=s')\frac{K_M(i,s)}{M}\right)(Z_i-\mu_Z(t_i^*,X_i))\right.\\
&+\frac{1}{n}\sum_i\1(t_i^*=s)(\mu_Z(s,X_i)-\mu_Z(s',X_i))\\
&+\left.\frac{1}{n}\sum_i\left(\1(t_i^*=s)-\1(t_i^*=s')\frac{K_M(i,s)}{M}\right)\mu_Z(s',X_i)\right\}\\
:&=\frac{n}{N_s}\left(T_{1,n}+T_{2,n}+T_{3,n}\right).
\end{align*}
We look at these three terms separately. First, $\E[T_{1,n}|\tstar,\X]=0$ and
\begin{align*}
\V[T_{1,n}|\tstar,\X]&=\frac{1}{n^2}\sum_i \left(\1(t_i^*=s)-\1(t_i^*=s')\frac{K_M(i,s)}{M}\right)^2\sigma^2_Z(t_i^*,X_i)
\end{align*}
from which 
\begin{align*}
\V[T_{1,n}]&=\frac{1}{n}\E\left[\left(\1(t_i^*=s)-\1(t_i^*=s')\frac{K_M(i,s)}{M}\right)^2\sigma^2_Z(t_i^*,X_i)\right]\\
&\le \frac{C}{n}\left(1+\frac{1}{M^2}\E\left[K_M(i,s)^2\right]\right)\\
&\to 0
\end{align*}
because by Lemma 3 in \cite{Abadie-Imbens_2006_ECMA}, $\E\left[K_M(i,s)^2\right]$ is uniformly bounded. Thus, $T_{1,n}=o_\P(1)$. Next,
\begin{align*}
T_{2,n}&=\frac{1}{n}\sum_i\1(t_i^*=s)(\mu_Z(s,X_i)-\mu_Z(s',X_i))\\
&\to_\P \E\left[\mu_Z(s,X_i)-\mu_Z(s',X_i)|t_i^*=s\right]p_s\\
&=\left(\E[Z_i|t_i^*=s]-\E[\mu_Z(s',X_i)|t_i^*=s]\right)p_s.
\end{align*}
Finally, for the third term,
\begin{align*}
T_{3,n}&=\frac{1}{n}\sum_i\left(\1(t_i^*=s)-\1(t_i^*=s')\frac{K_M(i,s)}{M}\right)\mu_Z(s',X_i)\\
&=\frac{1}{n}\sum_i\1(t_i^*=s)\frac{1}{M}\sum_{m=1}^M(\mu_Z(s',X_i)-\mu_Z(s',X_{j_m(i,s')}))
\end{align*}
By a Taylor expansion,
\begin{align*}
T_{3,n}&=\frac{1}{n}\sum_i\1(t_i^*=s)\frac{1}{M}\sum_{m=1}^M\dot{\mu}_Z(s',\tilde{X})\norm{X_i-X_{j_m(i,s')}}
\end{align*}
where $\dot\mu_Z(s,x)=\partial \mu_Z(s,x)/\partial x$ and $\tilde{X}$ is a (random) intermediate point between $X_i$ and $X_{j_m(i)}$. Note that by continuity of $\mu_Z(s,x)$ and compactness, the derivative is uniformly bounded and thus 
\begin{align*}
T_{3,n}&\le \frac{1}{n}\sum_i\1(t_i^*=s)\frac{C}{M}\sum_{m=1}^M\norm{X_i-X_{j_m(i,s')}}\le \frac{C}{n}\sum_i\norm{X_i-X_{j_M(i,s')}}.
\end{align*}
Finally, note that
\begin{align*}
\E\left[\left(\frac{1}{n}\sum_i\norm{X_i-X_{j_M(i,s')}}\right)^2\right]&=\frac{1}{n^2}\sum_i\E\left[\norm{X_i-X_{j_M(i,s')}}^2\right]\\
&+\frac{2}{n^2}\sum_i\sum_{l>i}\E\left[\norm{X_i-X_{j_M(i,s')}}\norm{X_l-X_{j_M(l,s')}}\right]\\
&=\frac{1}{n}\E\left[\norm{X_i-X_{j_M(i,s')}}^2\right]+\frac{n(n-1)}{n^2}\E\left[\norm{X_i-X_{j_M(i,s')}}\norm{X_l-X_{j_M(l)}}\right]\\
&\le \E\left[\norm{X_i-X_{j_M(i,s')}}^2\right]
\end{align*}
by identical distributions and Cauchy-Schwarz. Thus,
\[T_{3,n}\le \E\left[\norm{X_i-X_{j_M(i,s')}}^2\right]=\frac{1}{n^{1/q}}\E\left[n^{1/q}\norm{X_i-X_{j_M(i,s')}}^2\right]\to 0\]
by Lemma 2 in \citet{Abadie-Imbens_2006_ECMA}. Thus $T_{3,n}=o_\P(1)$. Therefore,
\begin{align*}
T_n&=\frac{n}{N_s}\left(T_{1,n}+T_{2,n}+T_{3,n}\right)\to_\P \E[Z_i|t_i^*=s]-\E[\mu_Z(s',X_i)|t_i^*=s]
\end{align*}
and
\begin{align*}
\frac{1}{N_s}\sum_i\1(t_i^*=s')\frac{K_M(i,s)}{M}Z_i&\to_\P \E[\mu_Z(s',X_i)|t_i^*=s]
\end{align*}
as required. For the second result,
Let $\X_{s'}=(X_i)_{i:t_i^*=s'}$ be the subset of covariates in the comparison cohort $t_i^*=s'$ and let $m(s',x)=\E[Z_i^2|t_i^*=s',X_i=x]$. We have that
\begin{align*}
\frac{1}{N_s}\sum_i\1(t_i^*=s')\frac{K_M(i)^2}{M^2}Z_i^2&=\frac{n}{N_s}\cdot \frac{1}{M^2}\cdot T_n,\quad T_n=\frac{1}{n}\sum_i\1(t_i^*=s')K_M(i,s)^2Z_i^2
\end{align*}
and
\begin{align*}
\E[T_n]&=\E[\1(t_i^*=s')K_M(i,s)^2Z_i^2]=\E\left[\1(t_i^*=s')K_M(i,s)^2m(s',X_i)\right]\\
&=\E\left[\left.\E[K_M(i,s)^2|\tstar,\X_{s'},t_i^*=s']m(s',X_i)\right\vert t_i^*=s'\right]p_{s'}
\end{align*}
Now,
\begin{align*}
K_M(i,s)|\tstar,\X_{s'},t_i^*=s'\sim Binomial(N_s,\nu_s(\mathcal{A}_M(X_i)))
\end{align*}
where $\nu_s(\cdot)$ is the conditional distribution of $X|t^*=s$ and $\mathcal{A}_M(x)$ is the catchment area, see \citet{Abadie-Imbens_2006_ECMA} and \citet{Chen-Han_2024_wp} for details. This implies that
\[\E[K_M(i,s)^2|\tstar,\X_{s'},t_i^*=s']=N_s\nu_s(\mathcal{A}_M(X_i))+N_s(N_s-1)\nu_s(\mathcal{A}_M(X_i))^2\]
and thus
\begin{align*}
\E\left[K_M(i,s)^2m(s',X_i)|t_i^*=s'\right]&=\E\left[N_s\nu_s(\mathcal{A}_M(X_i))m(s',X_i)|t_i^*=s'\right]\\
&+\E\left[N_s(N_s-1)\nu_s(\mathcal{A}_M(X_i))^2m(s',X_i)|t_i^*=s'\right].
\end{align*}
For the first term, note that:
\begin{align*}
\E\left[N_s\nu_s(\mathcal{A}_M(X_i))m(s',X_i)|t_i^*=s'\right]&=\E\left[\frac{N_s}{N_{s'}}N_{s'}\E[\nu_s(\mathcal{A}_M(X_i))|t_i^*=s',X_i]m(s',X_i)|t_i^*=s'\right]\\
&=\frac{p_s}{p_{s'}}\E\left[N_{s'}\E[\nu_s(\mathcal{A}_M(X_i))|t_i^*=s',X_i]m(s',X_i)|t_i^*=s'\right]\\
&+\E\left[\left(\frac{N_s}{N_{s'}}-\frac{p_s}{p_{s'}}\right)N_{s'}\E[\nu_s(\mathcal{A}_M(X_i))|t_i^*=s',X_i]m(s',X_i)|t_i^*=s'\right].
\end{align*}
By Assumption \ref{assu:reg}(1) and Lemma 5.3 in \citet{Chen-Han_2024_wp}, $N_{s'}\E[\nu_s(\mathcal{A}_M(X_i))|t_i^*=s',X_i]m(s',X_i)$ is uniformly bounded. Then,
\begin{align*}
\E\left[\left(\frac{N_s}{N_{s'}}-\frac{p_s}{p_{s'}}\right)N_{s'}\E[\nu_s(\mathcal{A}_M(X_i))|t_i^*=s',X_i]m(s',X_i)|t_i^*=s'\right]&\le C \E\left[\abs{\frac{N_s}{N_{s'}}-\frac{p_s}{p_{s'}}}\right].
\end{align*}
Now,
\begin{align*}
\E\left[\abs{\frac{N_s}{N_{s'}}-\frac{p_s}{p_{s'}}}\right]&=\E\left[\abs{\frac{N_s}{N_{s'}}-\frac{p_s}{p_{s'}}}\1\left(\abs{\frac{N_s}{N_{s'}}-\frac{p_s}{p_{s'}}}>\delta\right)\right]\\
&+\E\left[\abs{\frac{N_s}{N_{s'}}-\frac{p_s}{p_{s'}}}\1\left(\abs{\frac{N_s}{N_{s'}}-\frac{p_s}{p_{s'}}}\le \delta\right)\right]\\
&\le \E\left[\abs{\frac{N_s}{N_{s'}}-\frac{p_s}{p_{s'}}}\1\left(\abs{\frac{N_s}{N_{s'}}-\frac{p_s}{p_{s'}}}>\delta\right)\right]+\delta\\
&\le \E\left[\abs{\frac{N_s}{N_{s'}}-\frac{p_s}{p_{s'}}}^2\right]^{1/2}\P\left[\abs{\frac{N_s}{N_{s'}}-\frac{p_s}{p_{s'}}}>\delta\right]^{1/2}+\delta\\
\to\delta 
\end{align*}
because $N_s/N_{s'}\to_\P p_s/p_{s'}$ and $\E[\abs{N_s/N_{s'}-p_s/p_{s'}}^2]$  is bounded by Lemma S3 in \citet{Abadie-Imbens_2016_ECMA}. Because $\delta$ is arbitrary, it follows that:
\[\E\left[\left(\frac{N_s}{N_{s'}}-\frac{p_s}{p_{s'}}\right)N_{s'}\E[\nu_s(\mathcal{A}_M(X_i))|t_i^*=s',X_i]m(s',X_i)|t_i^*=s'\right]\to0\]
and thus
\begin{align*}
\E\left[N_s\nu_s(\mathcal{A}_M(X_i))\sigma^2(s',X_i)|t_i^*=s'\right]&=\frac{p_s}{p_{s'}}\E\left[N_{s'}\E[\nu_s(\mathcal{A}_M(X_i))|t_i^*=s',X_i]m(s',X_i)|t_i^*=s'\right]+o(1)\\
&\to \frac{p_s}{p_{s'}}\E\left[\left.M\frac{f_s(X_i)}{f_{s'}(X_i)}m(s',X_i)\right\vert t_i^*=s'\right]
\end{align*}
by Theorem 4.1 and Lemma 5.3 in \citet{Chen-Han_2024_wp} and the dominated convergence theorem. Also note that:
\begin{align*}
\frac{f_s(X_i)}{f_{s'}(X_i)}&=\frac{e_s(X_i)/p_s}{e_{s'}(X_i)/p_{s'}}
\end{align*}
from which
\begin{align*}
\E\left[N_s\nu_s(\mathcal{A}_M(X_i))m(s',X_i)|t_i^*=s'\right]&\to_\P M\E\left[\left.\frac{e_s(X_i)}{e_{s'}(X_i)}m(s',X_i)\right\vert t_i^*=s'\right]\\
&=\frac{M}{p_{s'}}\E[e_s(X_i)m(s',X_i)].
\end{align*}
Similarly, for the second term,
\begin{align*}
\E\left[N_s(N_s-1)\nu_s(\mathcal{A}_M(X_i))^2m(s',X_i)|t_i^*=s'\right]=\left(\frac{p_s}{p_{s'}}\right)^2\E\left[N_{s'}^2\E[\nu_s(\mathcal{A}_M(X_i))^2|t_i^*=s',X_i]m(s',X_i)|t_i^*=s'\right]\\
+\E\left[\left(\frac{N_s(N_s-1)}{N_{s'}^2}-\left(\frac{p_s}{p_{s'}}\right)^2\right)N_{s'}^2\E[\nu_s(\mathcal{A}_M(X_i))^2|t_i^*=s',X_i]m(s',X_i)|t_i^*=s'\right].
\end{align*}
By Theorem 4.1 and Lemma 5.3 in \citet{Chen-Han_2024_wp}, 
\begin{align*}
\E\left[N_{s'}^2\E[\nu_s(\mathcal{A}_M(X_i))^2|t_i^*=s',X_i]m(s',X_i)|t_i^*=s'\right]&\to \alpha(M,q) \E\left[\left.\left(\frac{f_s(X_i)}{f_{s'}(X_i)}\right)^2m(s',X_i)\right\vert t_i^*=s'\right]\\
&=\alpha(M,q)\left(\frac{p_{s'}}{p_s^2}\right)\E\left[\frac{e_s(X_i)^2}{e_{s'}(X_i)}m(s',X_i)\right]
\end{align*}
On the other hand, by previous arguments
\[\E\left[\left(\frac{N_s(N_s-1)}{N_{s'}^2}-\left(\frac{p_s}{p_{s'}}\right)^2\right)N_{s'}^2\E[\nu_s(\mathcal{A}_M(X_i))^2|t_i^*=s',X_i]m(s',X_i)|t_i^*=s'\right]\to0.\]
Thus
\begin{align*}
\E\left[N_s(N_s-1)\nu_s(\mathcal{A}_M(X_i))^2m(s',X_i)|t_i^*=s'\right]&\to_\P\frac{\alpha(M,q)}{p_{s'}}\E\left[\frac{e_s(X_i)^2}{e_{s'}(X_i)}m(s',X_i)\right].
\end{align*}
and therefore
\begin{align*}
\E\left[K_M(i,s)^2m(s',X_i)|t_i^*=s'\right]&\to_\P\frac{M}{p_{s'}}\E[e_s(X_i)m(s',X_i)]+\frac{\alpha(M,q)}{p_{s'}}\E\left[\frac{e_s(X_i)^2}{e_{s'}(X_i)}m(s',X_i)\right].
\end{align*}
Thus,
\begin{align*}
\E[T_n]&\to_\P M\E[e_s(X_i)m(s',X_i)]+\alpha(M,q)\E\left[\frac{e_s(X_i)^2}{e_{s'}(X_i)}m(s',X_i)\right].
\end{align*}
Finally, note that:
\begin{align*}
\V[T_n]&=\frac{1}{n}\V\left[\1(t_i^*=s')K_M(i,s)^2m(s',X_i)\right]+\frac{1}{n}\E\left[\1(t_i^*=s')K_M(i,s)^4\V[Z_i^2|t_i^*=s',X_i]\right]\\
&+\frac{2}{n^2}\sum_i\sum_{j>i}\cov\left(\1(t_i^*=s')K_M(i,s)^2m(s',X_i),\1(t_j^*=s')K_M(j,s)^2m(s',X_j)\right)\\
&\le \frac{1}{n}\V\left[\1(t_i^*=s')K_M(i,s)^2m(s',X_i)\right]+\frac{1}{n}\E\left[\1(t_i^*=s')K_M(i,s)^4\V[Z_i^2|t_i^*=s',X_i]\right]\\
&\le \frac{C}{n}\E\left[K_M(i,s)^4\right]\\
&\to 0
\end{align*}
where the first inequality follows from the fact that $K_M(i,s)$ and $K_M(j,s)$ are negatively correlated \citep[][p.40]{Abadie-Imbens_2002_NBER}, the second inequality follows because the conditional variance is uniformly bounded and the convergence follows from the fact that $\E[K_M(i,s)^4]$ is bounded by Lemma 3 in \citet{Abadie-Imbens_2006_ECMA}. This implies that
\[T_n\to_\P  M\E[e_s(X_i)m(s',X_i)]+\alpha(M,q)\E\left[\frac{e_s(X_i)^2}{e_{s'}(X_i)}m(s',X_i)\right]\]
and thus
\begin{align*}
\frac{1}{N_s}\sum_i\1(t_i^*=s')\frac{K_M(i,s)^2}{M^2}Z_i^2&=\frac{n}{N_s}\cdot \frac{1}{M^2}\cdot T_n\\
&\to_\P \frac{1}{M}\E\left[e_s(X_i)\frac{m(s',X_i)}{p_s}\right]+\frac{\alpha(M,q)}{M^2}\E\left[\frac{e_s(X_i)^2}{e_{s'}(X_i)}\frac{m(s',X_i)}{p_s}\right]
\end{align*}
which completes the proof. $\square$


\end{document}